\documentclass[journal=jacsat,manuscript=article]{achemso}
\usepackage{chemformula} 
\usepackage[T1]{fontenc} 
\usepackage{times}

\usepackage{amssymb} 
\usepackage{times}

\usepackage{mciteplus}
\mciteErrorOnUnknownfalse

\newcommand{\be}{\begin{equation}}
\newcommand{\ee}{\end{equation}}
\newcommand{\bea}{\begin{eqnarray}}
\newcommand{\eea}{\end{eqnarray}}
\newcommand{\beas}[1]{\begin{subequations}\label{#1}\bea}
\newcommand{\eeas}{\eea\end{subequations}}
\newcommand{\kB}{ k_\text{B} }

\newcommand{\figref}[1]{{\color{orange}\ref{#1}}}
\newcommand{\newzeta}{\xi}

\SectionNumbersOn

\author{Tianchi Li}
\affiliation[A]{Soft and Living Materials, Department of Materials, ETH Zurich, CH--8093 Zurich, Switzerland}
\affiliation[A]{Soft and Living Materials, Department of Materials, ETH Zurich, CH--8093 Zurich, Switzerland}
\author{Eric R. Dufresne}
\affiliation[A]{Soft and Living Materials, Department of Materials, ETH Zurich, CH--8093 Zurich, Switzerland}
\author{Martin Kr\"oger}
\affiliation{Polymer Physics, Department of Materials, ETH Zurich, CH--8093 Zurich, Switzerland}
\altaffiliation{Magnetism and Interface Physics, Department of Materials, ETH Zurich, CH--8093 Zurich, Switzerland}
\author{Stefanie Heyden}
\affiliation[A]{Soft and Living Materials, Department of Materials, ETH Zurich, CH--8093 Zurich, Switzerland}
\email{stefanie.heyden@mat.ethz.ch}

\title[Siloxane molecules: Nonlinear elastic behavior and fracture characteristics]
  {Siloxane molecules: Nonlinear elastic behavior and fracture characteristics}


\begin{document}

\begin{tocentry}

\includegraphics[width=8.4cm]{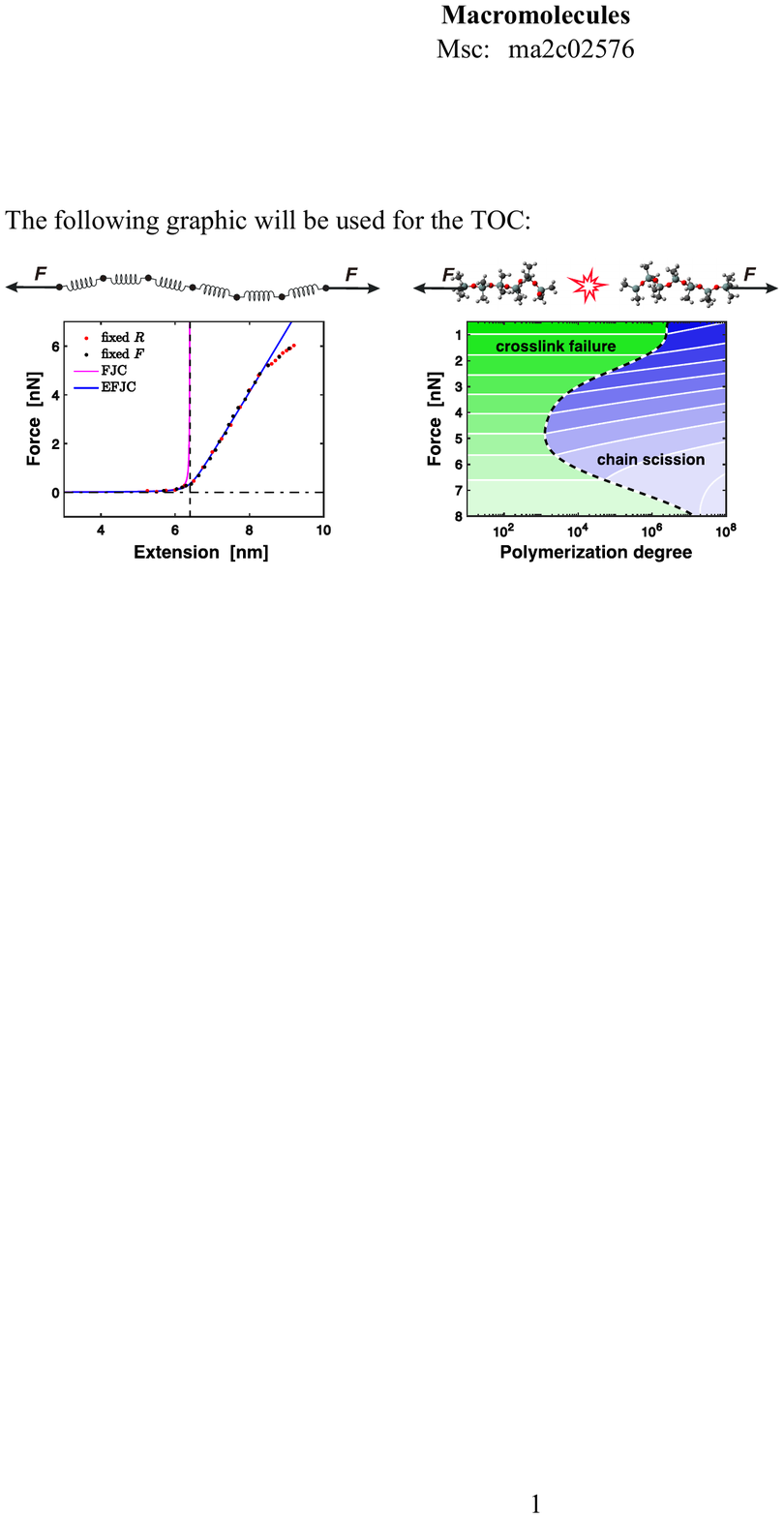}

\end{tocentry}

\begin{abstract}
Fracture phenomena in soft materials span multiple length- and timescales. This poses a major challenge in computational modeling and predictive materials design. To pass quantitatively from molecular- to continuum scales, a precise representation of the material response at the molecular level is vital.
Here, we derive the nonlinear elastic response and fracture characteristics of individual siloxane molecules using molecular dynamics (MD) studies.
For short chains, we find deviations from classical scalings for both the effective stiffness and mean chain rupture times.
A simple model of a non-uniform chain of Kuhn segments captures the observed effect and agrees well with MD data.
We find that the dominating fracture mechanism depends on the applied force scale in a non-monotonic fashion.
This analysis suggests that common polydimethylsiloxane (PDMS) networks fail at crosslinking points.
Our results can be readily lumped into coarse-grained models. Although focusing on PDMS as a model system, our study presents a general procedure to pass beyond the window of accessible rupture times in MD studies employing mean first passage time theory, which can be exploited for arbitrary molecular systems.
\end{abstract}

\section{Introduction}

Most things in life start small.
This basic concept also applies to the failure of soft materials, emerging from the rupture of interatomic bonds.
Predicting the fracture journey that follows becomes a question of failure mechanisms and lengthscales \cite{Zhao:2014,Bai:2019,Long:2021}.
In view of failure mechanisms, Lake-Thomas theory has formed our understanding of how much energy it takes to break an elastic chain \cite{Lake:1967}. 
When a crack propagates within a stretched elastic material, each repeat unit within chains crossing the fracture plane stores energy.
The resultant fracture energy should thus reflect the elastic energy stored within the entire chain instead of pure single bond scission.
Recent works on tough hydrogels hint at a more complicated picture, in which network characteristics such as entanglements have a crucial effect on fracture \cite{Suo:2021,Suo:2022}.

Multiple lengthscales form the basis of the classical fracture mechanics picture. 
In the ideally brittle limit, dissipation and material failure occur on the scale of the atomistic separation length.
In soft tough materials, the characteristic lengthscale in the continuum limit is the so-called \emph{elasto-adhesive} length.
This lengthscale is typically microscopic and represents the region of nonlinear elastic deformation around a macroscopic crack tip \cite{Long:2021}.
It can be coupled to molecular failure processes at small scales \cite{Lake:1967,Martina:2006}, as well as energy dissipation at the mesoscale \cite{Brown:2007,Tanaka:2007,Zhang:2015} and macroscopic effects such as crack blunting \cite{Jagota:2003,Seitz:2009}. 
In a recent work, scale-free cavity growth at constant driving pressure was accessed in the mesoscopic region \cite{Kimeaaz0418}.
In this picture, no well defined crack tip exists and corresponding process zones for the calculation of fracture energies becomes obsolete.

Fracture in soft solids thus displays manifold characteristics that are deviating from classical theories. To get further insight into what governs these deviations, multiple length- and timescales need to be bridged, which poses a major computational challenge.
Here, we address this challenge by providing a detailed description of the nonlinear elastic response of molecular building blocks up to fracture.
These building blocks can then provide starting grounds for higher level coarse grained models.

Previous studies on the force-extension relation and fracture of individual molecules encompass both experimental- and computational investigations.
Experimental studies include atomic force microscopy (AFM) \cite{AFM,AFM_macromolecules_1,AFM_macromolecules_2,AFM_macromolecules_3}, optical tweezers \cite{optical_tweezers,EWLC_marko_siggia,AFM_OT_macromolecules}, as well as magnetic tweezers \cite{FJC_smith,magnetic_tweezers,AFM_MT}. 
Due to its large accessible force range up to $\mathcal{O}($nN$)$ and high resolution, AFM has been widely adopted \cite{Butt_2005}.
Investigations using AFM comprise a wide spectrum, ranging from proteins \cite{Rief97reversibleunfolding,Fisher_2000}, DNA \cite{Rief_1999,Bustamante_2000}, polysaccharides \cite{AFM_polysaccharides}, poly(ethylene glycol) \cite{Oesterhelt_1999} and poly(methacrylic acid) \cite{AFM_PMAA_macromolecules} to polydimethylsiloxane (PDMS) \cite{AFM_PDMS}.

Using computational methods, ab initio molecular dynamics (AIMD) \cite{AIMD,AIMD_macromolecules} simulations allow for an on-the-fly computation of electronic structures based on quantum mechanics. 
While bond fracture can be modeled in this setting, high computational costs limit AIMD studies to $\mathcal{O}($nm$)$ and $\mathcal{O}($ps$)$ \cite{AIMD_PDMS_single,AIMD_PDMS_multiple,AIMD_PDMS_HMDSO}.
At higher length- and timescales, steered molecular dynamics (MD) simulations have emerged as the primary method in studying the force-extension behavior of molecules \cite{LU1998,LU1999,ISRALEWITZ2001,SMD_macromolecules}.
Classical MD methods are amenable of treating system sizes of several hundreds of nanometers and time scales on the order of nanoseconds.
However, atomic interactions are typically modeled \emph{via} empirical interatomic potentials, which require a predefined atomic connectivity remaining unchanged throughout simulations, such that fracture of interatomic bonds cannot be described.
As an alternative, bond-order based force fields were developed to bridge the gap between ab-initio and empirical force fields. 
Here, we derive the quasi-static force-extension and rupture properties of single molecules up to $\mathcal{O}(100\,\text{nm})$ and $\mathcal{O}($ns$)$ by enriching all-atom steered molecular dynamics simulations with a bond-order based force field (\emph{ReaxFF}) \cite{ReaxFF,PDMS_ReaxFF,Newsome2012,Soria2017,Soria2018}, with the help of the LAMMPS software package.\cite{PLIMPTON19951}
Unlike classical atomistic bond potentials, ReaxFF allows for different atomic bonding states, such that fracture events can be captured.
Simulations thus reduce the gap between length- and time scales accessible using ab initio computational methods and experimental approaches.
We focus on PDMS as a model system as used in previous studies \cite{Kimeaaz0418}, for which both linear PDMS and crosslinked PDMS are investigated.

\section{Molecular Dynamics Studies}

\subsection{Nonlinear elastic response}

\begin{figure*}[h]
\centering
    {\includegraphics[width=\textwidth]{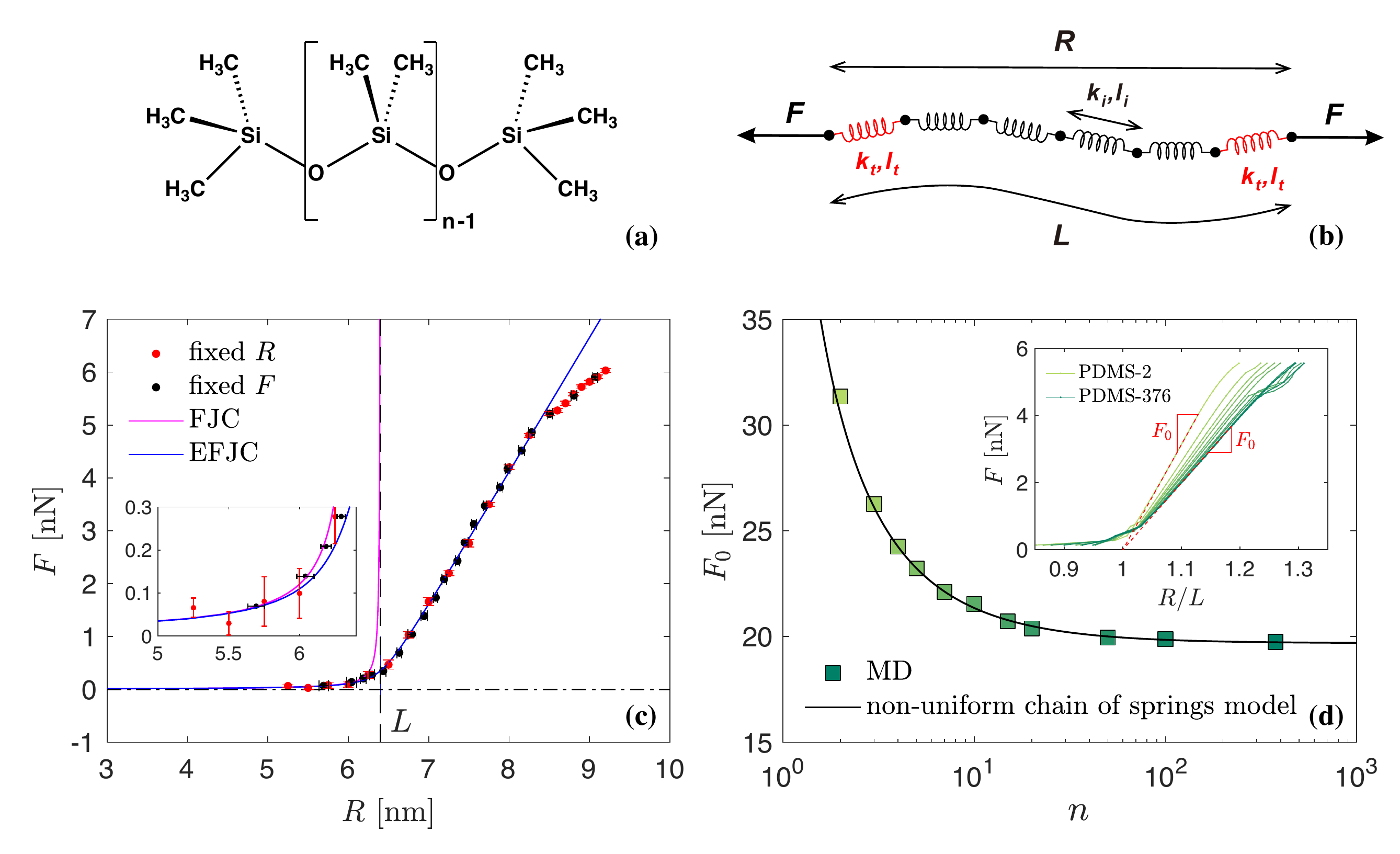}}
\caption{(\textbf{a}) Chemical structure of PDMS-$n$ illustrating $(n-1)$ repeating units.
(\textbf{b}) Non-uniform chain of springs model consisting of two types of Kuhn springs.
(\textbf{c}) Force-extension relation $F(R)$ of a single PDMS-27 molecule measured at $T=300$ K and $P=1$ atm in the presence of HMDSO solvent molecules. Red/black data points are obtained in a displacement/force controlled setting, respectively. 
Inset showing the small extension regime ($R<L$). 
The FJC model (pink solid upper line) and elastic FJC model (blue solid lower line) both capture the entropic regime $R\ll L$, while the latter shows better agreement in the enthalpic region $R>L$.
(\textbf{d}) Evolution of $F_0=L\, \partial F/\partial R$, obtained at $T=1$ K within the elastic regime as shown in the inset for $n=2$ up to $n=376$.
MD data is captured well by a model of non-uniform chain of springs, consisting of two types of Kuhn springs as sketched in (\textbf{b}).}
\label{fig1}
\end{figure*}

Prior to failure, the static molecular response is governed by entropic elasticity at extensions well below the unstretched contour length, and enthalpic elasticity at higher extensions.
The exact shape of this nonlinear elastic force-extension relation depends on the specific molecular structure under investigation.
To derive the nonlinear elastic response of siloxane molecules, PDMS-$n$ molecules of varying polymerization degree $n$ are created. 
Figure~\figref{fig1}a illustrates the chemical structure of PDMS-$n$ as an example. 
Each PDMS-$n$ molecule is embedded in a simulation box, which is set up both with and without solvent molecules.
When solvent molecules are present, periodic boundary conditions are applied.
We use hexamethyldisiloxane (HMDSO) molecules as a solvent, as interactions between HMDSO and PDMS do not alter the rupture behavior of PDMS (compared to interactions with itself) \cite{AIMD_PDMS_HMDSO}.
In comparison, trace amounts of water were found to lower the maximally attained rupture stretch \cite{AIMD_PDMS_HMDSO}.

All simulations are performed with the parameter set specifically trained and optimized for PDMS \cite{PDMS_ReaxFF}.
Without solvent molecules, the number of degrees of freedom is $3n_a$, with $n_a$ being the number of atoms in PDMS-$n$.
$n_a$ scales linearly with polymerization degree $n$.
For simulations in which solvent molecules are present, the overall system size increases by $3n_s$ degrees of freedom based on $n_s$ solvent molecules. There is no upper constraint on $n_s$. Its lower bound is set by the requirement of generating sufficiently large RVE's for subsequent steered molecular dynamics runs, preventing self-interactions.  
For simulations of PDMS-27 up to fracture, $n_s/n_a\sim 60$.

amounts to $60\times 3n_a$ degrees of freedom.
Here, the multiplicative factor of $60$ stems from the presence of solvent molecules.
In all simulations, a timestep of $0.1$\,fs is applied.
Systems are relaxed in an NPT ensemble at ambient conditions ($T=300$\,K, $p=1$\,atm).
Following relaxation, a constant repulsive force $F$ between the 2 terminal Si atoms is applied in an NVT ensemble.
In this ensemble, volume $V$ is held constant, such that the equilibrated end-to-end distance is purely based on the applied force (rescalings of the simulation box are prohibited).
Results are compared to a displacement controlled setting, in which both terminal Si atoms are held constant at fixed end-to-end distance $R$ and the exerted force is recorded.
Figure~\figref{fig1}c shows the nonlinear elastic reponse of PDMS-27 up to fracture.
The choice of boundary condition does not influence the force-extension relation in both entropic ($R\ll L$) and enthalpic ($R>L$) regimes, where the unstretched contour length $L$ marks the crossover point.
For comparison with classical polymer models, the inset of Figure~\figref{fig1}c highlights the divergence of the freely jointed chain model (FJC) \cite{FJC_smith} for an end-to-end chain distance $R$ approaching the unstretched contour length $L=6.4\,$nm.
In contrast, the elastic freely jointed chain model (EFJC) \cite{EFJC_smith} captures the nonlinear elastic force-extension relation also within the enthalpic regime $R>L$.
The change of slope at large forces is encoded in the $F$-dependent bond potential of mean force $U(b;F)$, as investigated in more detail in Section~\ref{sec:fracture}.

The force-extension curve at large deformation is linear. 
As expected from the EFJC model, the force-extension relation of a single polymer chain is given as 
\be 
    R(F) = L \left(1+\frac{F}{F_0}\right) {\cal L}(\newzeta), \qquad \text{where} \;\;
    \newzeta = \frac{F L_k}{\kB T} , \quad F_0 = K L \label{KL1}.
\ee
Equation~\eqref{KL1} represents a classical FJC model with an added elastic extension $(1+F/F_0)$. 
Within the entropic regime, elasticity is modeled via the Langevin function ${\cal L}(\newzeta)=\coth(\newzeta)-\newzeta^{-1}$.  
With the added elastic extension, the EFJC model introduces the effective Hookean spring constant $K$ as an additional elastic parameter within the enthalpic regime.

For computational efficiency, solvent molecules are removed for the determination of $F_0$, and $T=1$\,K is chosen to reduce thermal noise.
All other simulations are performed at $T=300$\,K.
Note that in this study, we focus on the enthalpic regime, in which temperature effects on the mechanical response become negligible.
Insensitivity of $F_0$ towards both temperature and solvent molecules in the enthalpic regime is tested for short oligomers (see Figures~\figref{fig-s2} and~\figref{fig-s4} in the Appendix).
We find that both temperature and solvent molecules do not influence $F_0$ in the ethalpic regime.
For small forces $F\ll F_0$ in the entropic limit, we have $R(F)\simeq F/\tilde{K}$, with $\tilde{K}=3k_BT/(LL_k)$ the elastic Hookean spring constant within the entropic regime.
For large forces $F\gg F_0$ in the enthalpic limit, $R(F)\simeq F/K$, as the Langevin function ${\cal L}(\newzeta)$ approaches unity for $\newzeta\gg 1$.
 $L_k$ is obtained from fitting the EFJC model to the measured force-extension curve at $T=300$ K, see Figure\ \figref{fig1}c.
A comparison to other hypothetical Kuhn segments (which consistently overpredict the unstretched contour length $L$) is shown in Figure~\figref{fig:Kuhn_length} in the Appendix).
Further support giving an independent estimate of $L_k$ from analyzing the Si-Si vector correlation function is given in supplementary Section~\ref{appendix-Kuhn}.
Here, $L_k=5.5\pm 0.7$\,\AA{} is the Kuhn length of the polymer chain corresponding to the mean end-to-end distance of a Si-O-Si-O-Si triplet, which in the following is abbreviated as Si-Si-Si.

For a low polymerization degree of $n=2$, the resultant slope $F_0$ when plotting force $F$ versus stretch $R/L$ attains its maximum value $F_0=31.35\,$nN (see inset in Figure~\ref{fig1}d).
$F_0$ decreases with increasing $n$ in a nonlinear fashion as illustrated in Figure~\figref{fig1}d.
Characteristic forces are calculated for a large range of polymerization degrees $n\in[2,3,4,5,7,10,15,20,50,100,376]$.
This differs from a classical model of $n$ identical springs in series, for which $K \propto 1/n$ and $L \propto n$.
To determine the source of this deviation, we track bond length distributions at fixed repulsive force between terminal Si atoms.
We find that within the enthalpic regime, internal triplet distances are shorter than terminal ones.
This difference in triplet distance distributions can be related to restrictions in bond angles and dihedrals at terminal atoms.
Endowing internal Si-Si-Si Kuhn segments with spring stiffness $k_i$ and equilibrium length $l_i$, whereas terminal Kuhn segments possess spring stiffness $k_t$ and equilibrium length $l_t$, gives the overall Hookean spring constant $K$ and contour length $L$ as
\be 
    K = \left(\frac{2}{k_t} + \frac{n/2-2}{k_i}\right)^{-1} , \qquad
    L = 2 l_t + \left(\frac{n}{2}-2\right) l_i  
\label{KL}.
\ee
Here, $n/2$ denotes the total number of Kuhn segments, where $n$ is the polymerization degree.
$k_t=47.3\,\pm 0.1$ nN/nm and $l_t= 0.52\pm 0.01\,$nm are directly computed from the bond length distribution of PDMS-4, which only consists of two terminal Kuhn segments ($k_t=2K$ and $l_t=L/2$).
Using a fit to simulation results for PDMS-5 to calculate the remaining free parameters, we find that $k_i=40.82\pm 0.01\,$nN/nm and $l_i=0.48\pm 0.01\,$nm of the internal Kuhn segment.
As shown by the solid black line in Figure~\figref{fig1}d, this model of a non-uniform chain of springs agrees well with MD data and captures the effect of terminal springs at small $n$, as well as convergence of $F_0$ to a plateau at large $n$.

To summarize, the elastic response of PDMS oligomers (both within entropic and enthalpic regimes) is characterized by the Si triplet length $L_k$ based on two reasons:
First, the entropic part of the force-extension curve suggests $L_k$ to be identical to the extension of a Si-Si-Si triplet.
Second, $n$-dependencies of $K$ and $L$ are consistently captured 
only if the number ($n/2$) of Si-Si-Si triplets is used in Equation\ \eqref{KL}.
In sharp contrast, the fracture behavior to be discussed next will be dominated by the $F$-dependent characteristics of single covalent atomic bonds.

\subsection{Fracture characteristics}
\label{sec:fracture}

At the molecular scale, fracture is stochastic.
An intuitive question to ask is 'Where and when does a network tend to break?'.
Here, we try to quantitatively answer this question for PDMS in terms of mean rupture times and preferred fracture modes.

To distinguish between rupture of the PDMS-$n$ backbone (\emph{chain scission}) and rupture at crosslinking sites (\emph{crosslink failure}), we take into account two different structures:
PDMS-$n$ as used in the previous Section, as well as two PDMS-4 molecules linked via a crosslinking site -$CH_2$-$CH_2$-.
Chain scission thus stems from the rupture of Si-O bonds, while crosslink failure results from rupturing Si-C bonds.
The accessible window of mean rupture times in MD studies lies in the range of $10^{-2}-10^0$\,ns.
The upper limit is set by computational feasibility, while the lower limit depends on the molecular vibration frequency of the polymer chain below which inertial effects dominate the response. 

\begin{figure*}
\centering
    {\includegraphics[width=0.9\textwidth]{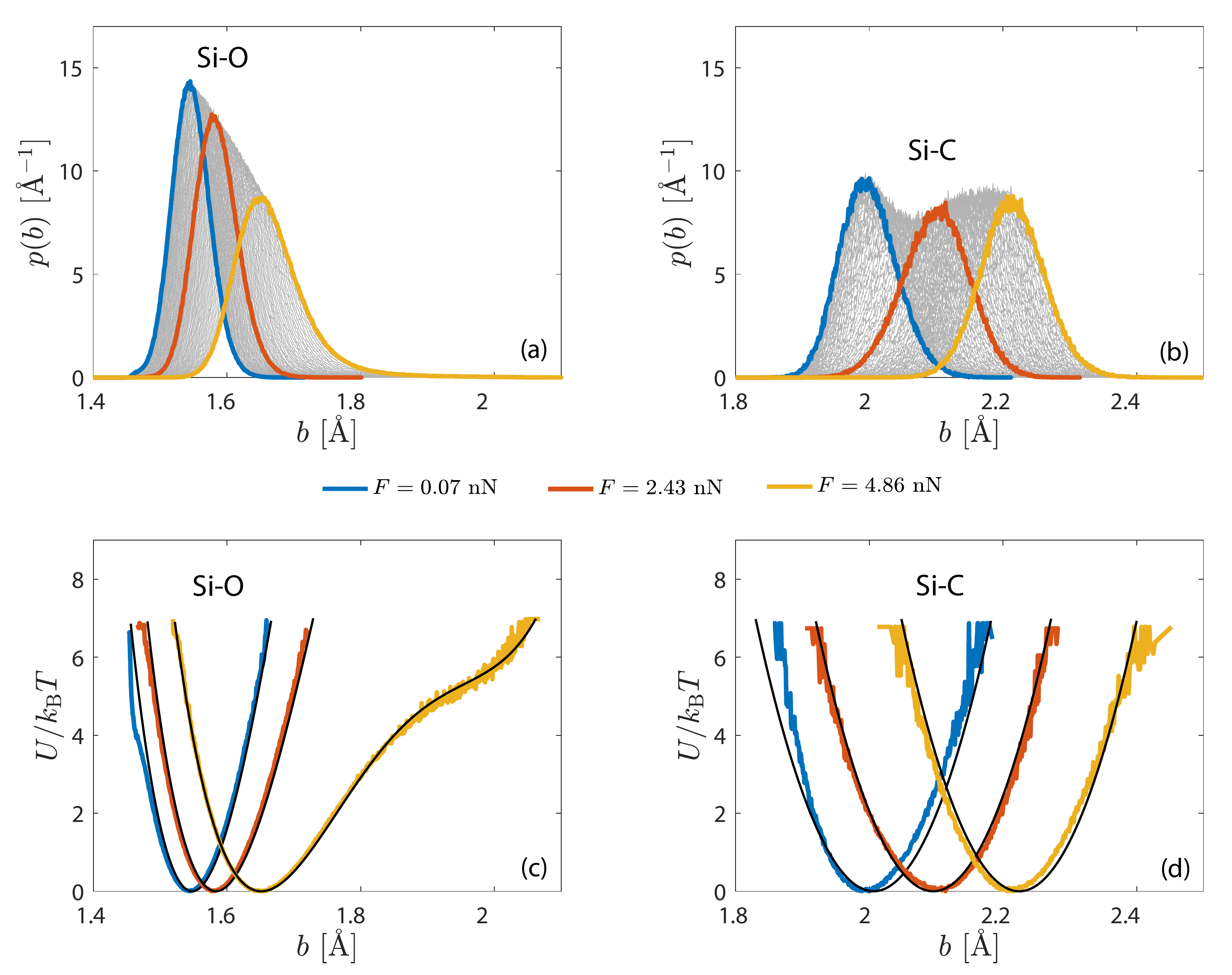}}
\caption{(\textbf{a and b}) Stationary nonequilibrium MD probability densities of Si-O and Si-C bond lengths on linked PDMS-4 at various force levels (simulations performed in the presence of solvent molecules).
Grey curves denote intermediate force levels.
 (\textbf{c and d}) Radial bond potential of mean force calculated from bond length distributions highlighted in (a,b). 
 Black curves represent polynomial fits of order 4 (Si-O) and 2 (Si-C).
 All simulations are performed at $T=300$ K.
 }
\label{fig2}
\end{figure*}

To extend beyond this rupture time window and determine mean bond rupture times $\tau(F)$ on longer timescales, we use a statistical extrapolation scheme.
Our approach renders a close analogy to the calculation of mean first passage times for chemical processes with a single reaction coordinate~\cite{Preston2021}, for the thermal or enforced breakage of discrete one-dimensional chains~\cite{Razbin2019}, Morse-chains~\cite{Puthur2002} and biomolecules~\cite{Berezhkovskii2019}.
Its derivation is provided in the Supplementary Information.
We proceed in the following way:
Stationary equilibrium Si-O and Si-C bond length probability densities $p(b)$ of PDMS-4, PDMS-376 and linked PDMS-4 are measured at different levels of constant force (\emph{cf.} Figure~\figref{fig2}). Using $p(b)$, we calculate mean chain rupture times, for which we need to pass from rupture times of single bonds to those of chains with $2n$ bonds.

At fixed force, directly measured single bond probability densities $p(b)$ serve to define an effective potential, with 
\be 
 p(b)=\frac{\exp(-U/\kB T)}{\int_0^{b_r} \exp(-U/\kB T)db}.
 \label{eq:bond_length_prob}
\ee 
Here, $b_r$ is the rupture bond length.
For later calculations of mean chain rupture times, we need a functional form of the effective bond potential $U$.
In the following, using the notation $U(b;F)$, we emphasize that the potential is a function of bond length $b$, while its parameters (in this case $b_1$) depend on $F$.
Numerically solving for $U$ from Equation~\eqref{eq:bond_length_prob} at different levels of applied force, we see that $U$ needs to satisfy the following properties:
It should have the generic form of a double-well potential (fourth order polynomial) with minima corresponding to two different equilibrium bond lengths (cf. Figure~\figref{fig:si_potential_derivs}).
Here, we denote $b_1$ as the equilibrium bond length corresponding to the first minimum, and choose $U(b_1)=0$ for convenience.
Most importantly, we observe a nearly $F$-independent parabolic shape of $U''(b)$ about its second minimum at $b\simeq b_2$, i.e., $U''(b)=c_2+k_2(b-b_2)^2$ (cf. Figure~\figref{fig:si_potential_derivs}).
Since $U''(b)$ exhibits a parabolic and $F$-independent shape and location, the corresponding parameters $b_2$, $c_2$, $k_2$ can be treated as $F$-independent constants.
This finding forms the basis of rendering our statistical extrapolation scheme feasible, since $b_1(F)$ remains as the only  force-dependent fitting parameter. 
In addition, we only observe a weak dependence of $b_1$ on $F$, which forms the basis for an extrapolation to higher force regimes. 
These observations lead to a functional form of $U(b;F)$ as
\begin{eqnarray} 
 U(b;F) &=& \int_{b_1}^b\int_{b_1}^x\left[ c_2 + k_2(y-b_2)^2 \right] \,dy\,dx
 \nonumber \\
 &=& \frac{(b-b_1)^2\{6c_2+k_2[3b_1^2+6b_2^2-8b_1 b_2+2(b_1-2b_2)b + b^2]\}}{12}.
 \label{eq:potential}
\end{eqnarray}

Bond length distributions (Figures~\figref{fig2}a and~\figref{fig2}b) are fitted to Equations\ \eqref{eq:bond_length_prob}, \eqref{eq:potential} with fitting parameters given in Table~\figref{tab_parameters}.
The resulting radial bond potentials of mean force are illustrated in Figure~\figref{fig2}c and~\figref{fig2}d, with corresponding polynomials of order 4 (Si-O bond) and order 2 (Si-C bond, for which $b_2=k_2=0$).
 Note that fitting deviations in Si-C potentials are based on restricting $b_1(F)$ to be the only force-dependent parameter.
 The largest possible instantaneous Si-O an Si-C bond length value in stable chain configurations is given in Table~\ref{tab_parameters}.
As shown in Figure~\figref{fig2}c, an increasing non-linearity develops with increasing tension for Si-O bonds.
This is rooted in bond angle potentials losing their dominance within the energy landscape due to the externally enforced alignment.
At this point, the remaining interactions (dihedral, Si-C, Si-H, O-H) come into play.

With an expression for $U(b;F)$ at hand, we proceed with the calculation of mean chain rupture times, passing from rupture times of single bonds to those of chains with $2n$ bonds. 
Furthermore, it needs to be verified that a theory neglecting inertia effects captures the attendant fracture characteristics.

Neglecting inertia effects, the mean rupture time of a single bond is calculated from the Fokker-Planck equation as \cite{Risken,Kampen,Gardiner} 
\be 
 \tau(F) = \frac{\zeta}{\kB T} \int_{b_1(F)}^{b_r} \int_0^z \frac{\Psi(y)}{\Psi(z)}\,dy\,dz ,
 \qquad \text{with} \quad \Psi(b) = \exp\left[-\frac{U(b;F)}{\kB T}\right].
\label{eq:meanfirstpassage}
\ee
This is a purely theoretical limit, as atomistic simulations (PDMS chains consist of multiple Si-O and Si-C bonds) measure $\tau_n$.
$\zeta$ is an a priori unknown friction coefficient, which will be determined later by matching theoretical rupture times $\tau_n(F)$ with those obtained from MD simulations.
Utilizing the Fokker-Planck approach, Equation~\eqref{FPbond} (or equivalently Brownian Dynamics simulations via Equation~\eqref{Langevinbond}) can be used to explore the rupture time distribution $p(t_r;F)$ of a single bond, which is nearly mono-exponential (apart from a small dip at $t_r\rightarrow 0$).

\begin{figure*}
\centering
    {\includegraphics[width=\textwidth]{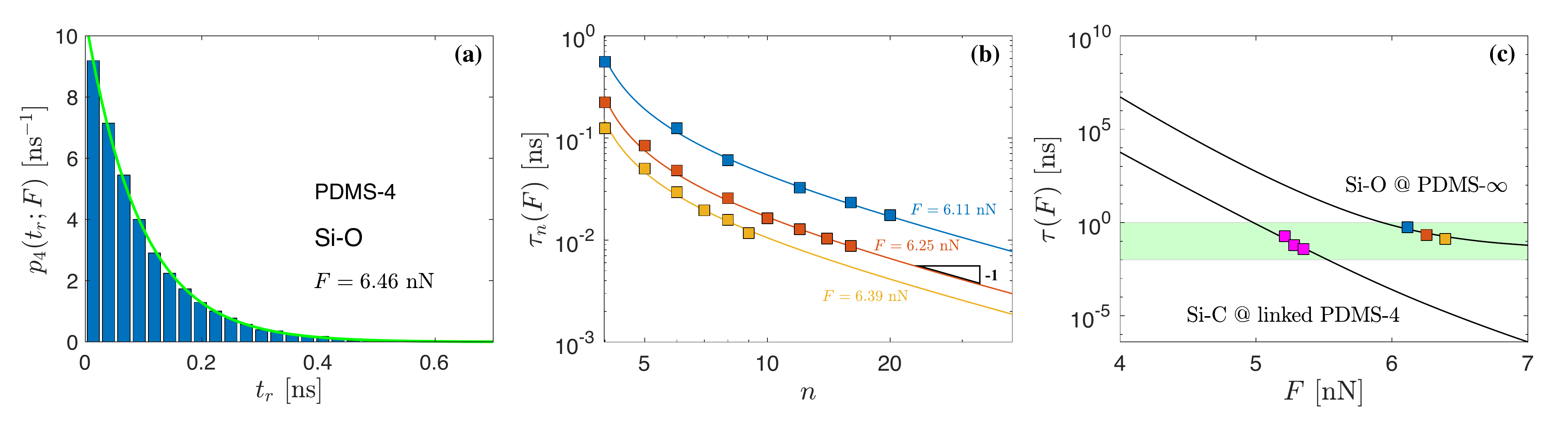}}
\caption{(\textbf{a}) Measured rupture time probability density $p_4(t_r;F)$ of PDMS-4 at constant stretching force $F=6.46$ nN. 
 Green solid line: Mono-exponential used in the calculation of mean rupture time $\tau_{n}(F)$.
 (\textbf{b}) Evolution of mean chain rupture time $\tau_n(F)$ with increasing polymerization degree $n$ at three different forces.
Each data point is the average of $10000$ simulations.
Solid lines are obtained from fitting the functional form given in Equation~\eqref{eq:3}.
For large $n$, $\tau_n(F) \propto 1/n$ as expected.  
Deviations for short chains are related to the effect of non-uniform chain of springs highlighted in Figure~\figref{fig1}d.
(\textbf{c}) Single-bond mean rupture time $\tau(F)$ for Si-O (obtained from \eqref{eq:3}) and Si-C (average of individual MD simulations). 
Solid lines: Solution to the mean first passage time problem~\eqref{eq:meanfirstpassage} based on $U$ (cf. Figure~\figref{fig2}b-c).
The shaded region highlights the rupture time window accessible in MD studies. All simulations are performed at $T=300$ K.
} 
\label{fig3}
\end{figure*}

In order to pass to the rupture time distribution $p_n(t_r;F)$ of a chain with polymerization degree $n$ (which thus contains $2n$ bonds), we assume independent bonds.
The probability of a chain (\emph{i.e.} at least one of its assumed identical bonds) rupturing during time interval $t$ after onset of $F$ at time $t=0$ is
\be 
 P_n(t;F) = 1 - \left[ 1 - \int_0^t p(t_r;F)dt_r\right]^{2n} = 1 - e^{-2nt/\tau(F)} . 
 \label{eq:rupture_prop}
\ee 
The term in parentheses in Equation~\eqref{eq:rupture_prop} denotes the probability of an individual bond staying intact until time $t$.
The probability distribution for rupture times $t_r$ of $n$-chains (PDMS-$n$) is thus $p_n(t_r;F)=(d/dt_r)P_n(t_r;F)=2ne^{-2nt_r/\tau(F)}/\tau(F)$, from which the mean chain rupture time $\tau_n(F)$ follows as $\tau_n(F)=\tau(F)/2n$.
We compare these theoretical expressions to those measured in MD simulations.
Figure~\ref{fig3}a shows measurements on PDMS-4. 
MD measurements show a mono-exponential shape of $p_4(t_r;F)$, which is in agreement with the Fokker-Planck prediction for a single bond and the assumption of independent bonds in chains of higher polymerization degree.

Figure~\figref{fig3}b illustrates $\tau_n(F)$ at three different constant stretching forces $F$, for which each data point is the average of $10000$ independent samples.
For large $n$, $\tau_n(F)$ approaches the expected $\propto 1/n$ limit.
Deviations from this scaling for short chains are reminiscent of the non-uniform chain of springs effect  highlighted in Figure~\figref{fig1}b.
Equivalent to the functional form given in Equation \eqref{KL}, we have
\be \label{eq:3}
    \frac{\tau_n(F)}{\textrm{ns}} = \frac{1}{a_0+(n-1)/a}, \qquad \frac{\tau(F)}{\textrm{ns}} = \lim_{n\rightarrow\infty} \frac{2 n\tau_n(F)}{\textrm{ns}} = 2a
\ee
Fitting parameters at $F=6.25$ nN are obtained as $a_0 = -24\pm 1$ and $a = 0.108\pm 0.004$, while at $F=6.11$ nN, $a_0 = -9.1\pm 0.4$ and $a = 0.28\pm 0.01$.
Single bond mean rupture times depicted in Figure~\figref{fig3}c are calculated as $\tau^{\rm Si-O}(F)=\lim_{n\rightarrow\infty}2n\tau_n(F)=2a$ ns.
For Si-C bonds (which are present twice in linked PDMS-4), force levels of $F=5.21\,$nN, $F=5.28\,$nN and $F=5.35\,$nN are investigated. 
This force range in MD already spans two decades in single-bond mean rupture time $\tau(F)$.
The resulting single bond mean rupture time is computed as $\tau^{\rm Si-C}(F)=2\tau^{\rm linked-PDMS-4}(F)$.

By matching the measured $\tau_n$ with the theoretically predicted one, we can furthermore determine $\zeta$ (which is the shape-preserving, force-independent vertical shift required to match measurement and theory). 
Solid lines in Figure~\figref{fig3}c illustrate the $\tau(F)$ resulting from the Fokker-Planck equation~\eqref{eq:meanfirstpassage}.
This solution allows to extend beyond the rupture time window accessible in MD studies (shaded region in Figure~\figref{fig3}c).
We find that the lifetime of a single representative Si-O bond is much longer than that of a Si-C bond, with an increasing gap for larger $F$ (based on the significant difference in potentials at high forces).

\begin{figure*}
\begin{tikzpicture}
    \node (schematic) at (0,0) {\includegraphics[width=\textwidth]{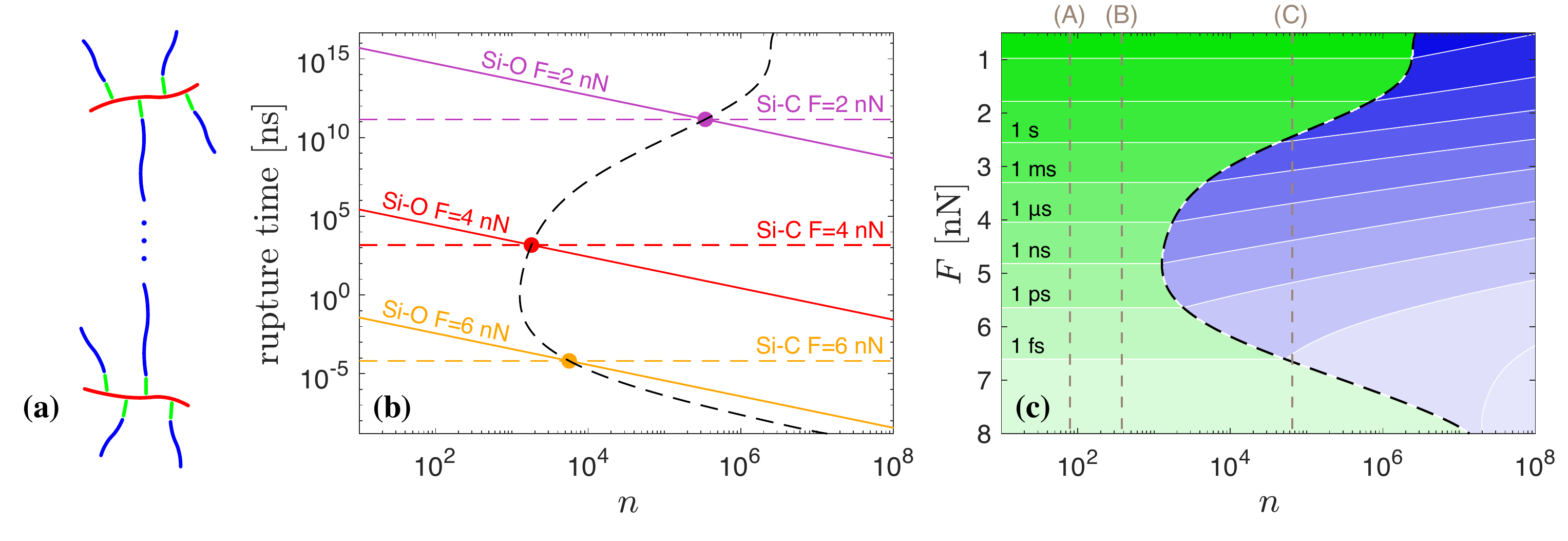}};
    \node[] at (6.7,0.2) {\small\color{gray}\text{\bf{chain scission}}};
    \node[] at (4,2) {\small\color{gray}\text{\bf{crosslink failure}}};
\end{tikzpicture}
\caption{Crosslink failure versus chain scission. 
(\textbf{a}) We model constituent molecules of monodisperse strands, testing for chain scission along the PDMS backbone (blue) versus failure at crosslinking junctions (green).
(\textbf{b}) Evolution of crossover polymerization degree $n_c$ (dashed black line) with increasing force $F$. 
Intersections of $\tau_n(F)$ for Si-O along the backbone and $\tau_2(F)$ for Si-C at two crosslinking junctions are illustrated at three different force levels.
(\textbf{c})
Effective rupture time of a network strand (contour lines) as a function of $n$ and $F$.
Green: Region dominated by failure of crosslinking junctions.
Blue: Chain scission dominated region. (A) \emph{Sylgard 184} ($n=78$), (B) \emph{DMS-V31} ($n=376$), (C) \emph{Sylgard 186} ($n=64848$) \cite{Sylgard-184/186}. All predictions apply to $T=300$ K.
}
\label{fig4}
\end{figure*}

To determine preferred fracture mechanisms, we note that with increasing polymerization degree $n$, $\tau_n^{\rm Si-O}(F)$ decreases, while $\tau_n^{\rm Si-C}(F)$ is constant (based on the constant number of Si-C bonds when focusing on the single chain level, see Figure~\figref{fig4}b).
Rupture of Si-O bonds (chain scission) and Si-C bonds (crosslink failure) thus becomes comparable at a crossover polymerization degree $n_c$.
Figure~\figref{fig4}b displays a comparison of mean chain rupture times $\tau_n(F)$ at different levels of applied force.
Tracking the crossover polymerization degree $n_c$ allows to identify two different failure regimes:
For $n<n_c$, crosslink failure is anticipated, while chain scission is the preferred failure mode for $n>n_c$.
Figure~\figref{fig4}c highlights the effective rupture time (taking into account both Si-O and Si-C bonds) as contour lines as a function of $n$ and $F$.
Again, the crossover polymerization degree $n_c$ differentiates a region dominated by crosslink failure (green) from a regime dominated by chain scission (blue).

We observe a strengthening effect in Si-O bonds with increasing force, which is reminiscent of phenomena observed in systems involving catch bonds, e.g., membrane-to-surface adhesion \cite{Danbo:1988}, myosin and actin \cite{Guo:2006}, or signaling receptors and their ligands \cite{Liu:2014,Reinherz:2016}.
This strengthening emerges as a 're-entrant' effect of crosslink failure for polymerization degrees $n>10^3$, which can be related to the higher order structure of Si-O bond potentials (see the nonlinearity developing at higher forces, Figures~\figref{fig2}, \figref{fig-s5} and~\figref{fig-4c}).
Si-O bonds are stable at low forces, at which the failure of crosslinking junctions is the dominating fracture mechanism.
At intermediate forces, chain scission dominates.
At high forces, at which the increasing stiffness of the second minimum in Si-O potentials comes into play, fracture characteristics are dominated by crosslink failure again.
With increasing $n$, the force regime dominated by chain scission grows, which is in keeping with Equation~\eqref{eq:rupture_prop}.

Typical siloxane materials used in the laboratory setting are highlighted in Figure~\figref{fig4}c in terms of polymerization degree $n$.
Single molecules in \emph{Sylgard 184} and \emph{DMS-V31} are entirely dominated by crosslink failure.
In contrast, individual molecules in \emph{Sylgard 186} feature a much higher polymerization degree.
As such, their fracture behavior strongly depends on the applied force, with crosslink failure in the low force regime transitioning to chain scission at $F>2.5\,$nN.

In the above, we model rupture time distributions using an inertia-free Fokker-Planck approach.
To justify this approach, it remains to investigate the absence of solvent molecules on rupture time distributions $p_n(t_r,F)$.
While the dynamics of bond lengths exhibits inertia effects, we find the rupture time distribution to be unaffected by the presence/absence of explicit solvent molecules, as shown in Figure~\figref{fig:rupture}.
Without solvent molecules, inertia effects are maximal and friction is absent.
The presence of solvent molecules provides additional noise and stochastic collisions, such that inertia effects are diminished.
The rupture time distribution is thus also unaffected by the degree of suppression of inertia effects.
Furthermore, we observe a nearly mono-exponential rupture time probability distribution, in which the majority of bonds does not break during the first oscillation but at later times.
This is in stark contrast to the dominance of inertial effects, for which bonds that are still intact after the first oscillation would never fail (the rupture criterion would not be fulfilled in subsequent oscillations if it was not already fulfilled in the first oscillation).

\section{Discussion and Outlook}

This work characterizes the nonlinear elastic- and fracture behavior of PDMS.
The nonlinear elastic repsonse of siloxane oligomers is captured well using the EFJC model, both within entropic- and enthalpic regimes.
At low polymerization degrees $n$, we find a deviation from a classical scaling of effective stiffness $K\propto1/n$, which can be captured with a simple model of non-uniform chains of springs, with each spring constituting a Kuhn segment.

Passing to the inelastic behavior of PDMS, we focus on rupture times of both Si-O bonds (present in the backbone of siloxane oligomers), as well as Si-C bonds (present at crosslinking sites). 
When calculating mean chain rupture times of siloxane oligomers, we again observe a deviation from classical scalings ($\tau_n(F)\propto1/n$) at low polymerization degrees. 
Similar to the nonlinear elastic part, a model including the 'end effect' on mean chain rupture times agrees well with simulations. 

We find that the lifetime of single Si-O bonds is much longer than that of a Si-C bond.
We define a crossover polymerization degree $n_c$ at which the dominating rupture mechanism of the crosslinked network passes from crosslink failure (rupture of Si-C bonds) to chain scission (rupture of Si-O bonds on PDMS-$n$).
To pass to long timescales, we use a statistical extrapolation scheme in order to understand bond fracture within the network. Surprisingly, we find that the dominating fracture mechanism depends on the applied force scale in a non-monotonous fashion.
The non-monotonic dependence of $n_c$ on $F$ is rooted in nonlinearities in the Si-O bond potential and the corresponding stiffness dependence on force at the two minima.

For single molecules in typical siloxane materials used in the laboratory setting, such as \emph{Sylgard 184} and \emph{DMS-V31}, our analysis suggests
breakage exclusively at crosslinks.
For individual molecules in \emph{Sylgard 186} featuring higher polymerization degrees however, the attendant failure mode depends on the magnitude of applied force.
While crosslink failure dominates at low forces, chain scission is expected for $F>2.5\,$nN.
It furthermore bears mentioning that our analysis focuses on the single chain level.
Possible inhomogeneities in force distribution based on network topology (and the resultant changes in fracture characteristics) furnish an important point for future studies. These could aid in elucidating network properties passing from the single chain to the continuum level, focusing on the influence of polymerization degree and crosslink density.

Our results provide building blocks, which can be readily used in coarse-grained higher scale models.
As an example, network models of nonlinear springs could be easily tailored to siloxane systems by implementing a material model corresponding to the nonlinear elastic response derived in this work.
In this setting, crosslinking molecules could be lumped into nodes representing crosslinking sites. 
Tracking attendant forces upon deformation would then allow for the implementation of fracture criteria corresponding to those derived in this work.

Finally, our work provides a generic procedure to pass beyond the window of accessible rupture times in MD studies, which can be applied for arbitrary molecular systems.

\begin{acknowledgement}
SH and TL gratefully acknowledge funding via the SNF Ambizione grant PZ00P2186041.
We are furthermore thankful for many insightful discussions with our Soft $\&$ Living Materials peers Dr. Robert Style and Dr. Nicolas Bain.
\end{acknowledgement}

\begin{suppinfo}

The following file is available free of charge.
\begin{itemize}
  \item SI.pdf: (i) Characteristic force at finite temperature, (ii) Kuhn length, (iii) parameters of the double well potential, (iv) mean bond rupture time, (v) influence of boundary conditions, (vi) bond length distributions and potentials, (vii) rupture times with and without HMDSO solvent, (viii) mean chain rupture times for different polymerization degrees.
\end{itemize}

\end{suppinfo}

\bibliography{refs}

\providecommand{\latin}[1]{#1}
\makeatletter
\providecommand{\doi}
  {\begingroup\let\do\@makeother\dospecials
  \catcode`\{=1 \catcode`\}=2 \doi@aux}
\providecommand{\doi@aux}[1]{\endgroup\texttt{#1}}
\makeatother
\providecommand*\mcitethebibliography{\thebibliography}
\csname @ifundefined\endcsname{endmcitethebibliography}
  {\let\endmcitethebibliography\endthebibliography}{}
\begin{mcitethebibliography}{62}
\providecommand*\natexlab[1]{#1}
\providecommand*\mciteSetBstSublistMode[1]{}
\providecommand*\mciteSetBstMaxWidthForm[2]{}
\providecommand*\mciteBstWouldAddEndPuncttrue
  {\def\EndOfBibitem{\unskip.}}
\providecommand*\mciteBstWouldAddEndPunctfalse
  {\let\EndOfBibitem\relax}
\providecommand*\mciteSetBstMidEndSepPunct[3]{}
\providecommand*\mciteSetBstSublistLabelBeginEnd[3]{}
\providecommand*\EndOfBibitem{}
\mciteSetBstSublistMode{f}
\mciteSetBstMaxWidthForm{subitem}{(\alph{mcitesubitemcount})}
\mciteSetBstSublistLabelBeginEnd
  {\mcitemaxwidthsubitemform\space}
  {\relax}
  {\relax}

\bibitem[Zhao(2014)]{Zhao:2014}
Zhao,~X. Multi-scale multi-mechanism design of tough hydrogels: building
  dissipation into stretchy networks. \emph{Soft Matter} \textbf{2014},
  \emph{10}, 672--687\relax
\mciteBstWouldAddEndPuncttrue
\mciteSetBstMidEndSepPunct{\mcitedefaultmidpunct}
{\mcitedefaultendpunct}{\mcitedefaultseppunct}\relax
\EndOfBibitem
\bibitem[Bai \latin{et~al.}(2019)Bai, Yang, and Suo]{Bai:2019}
Bai,~R.; Yang,~J.; Suo,~Z. Fatigue of hydrogels. \emph{Europ. J. Mech. A}
  \textbf{2019}, \emph{74}, 337--370\relax
\mciteBstWouldAddEndPuncttrue
\mciteSetBstMidEndSepPunct{\mcitedefaultmidpunct}
{\mcitedefaultendpunct}{\mcitedefaultseppunct}\relax
\EndOfBibitem
\bibitem[Long \latin{et~al.}(2021)Long, Hui, Gong, and Bouchbinder]{Long:2021}
Long,~R.; Hui,~C.-Y.; Gong,~J.; Bouchbinder,~E. The Fracture of Highly
  Deformable Soft Materials: A Tale of Two Length Scales. \emph{Annu. Rev.
  Condens. Matter Phys.} \textbf{2021}, \emph{12}, 71--94\relax
\mciteBstWouldAddEndPuncttrue
\mciteSetBstMidEndSepPunct{\mcitedefaultmidpunct}
{\mcitedefaultendpunct}{\mcitedefaultseppunct}\relax
\EndOfBibitem
\bibitem[Lake and Thomas(1967)Lake, and Thomas]{Lake:1967}
Lake,~G.; Thomas,~A. The strength of highly elastic materials. \emph{Proc. R.
  Soc. A} \textbf{1967}, \emph{300}, 108--119\relax
\mciteBstWouldAddEndPuncttrue
\mciteSetBstMidEndSepPunct{\mcitedefaultmidpunct}
{\mcitedefaultendpunct}{\mcitedefaultseppunct}\relax
\EndOfBibitem
\bibitem[Kim \latin{et~al.}(2021)Kim, Zhang, Shi, and Suo]{Suo:2021}
Kim,~J.; Zhang,~G.; Shi,~M.; Suo,~Z. Fracture, fatigue, and friction of
  polymers in which entanglements greatly outnumber cross-links. \emph{Science}
  \textbf{2021}, \emph{374}, 212--216\relax
\mciteBstWouldAddEndPuncttrue
\mciteSetBstMidEndSepPunct{\mcitedefaultmidpunct}
{\mcitedefaultendpunct}{\mcitedefaultseppunct}\relax
\EndOfBibitem
\bibitem[Nian \latin{et~al.}(2022)Nian, Kim, Bao, and Suo]{Suo:2022}
Nian,~G.; Kim,~J.; Bao,~X.; Suo,~Z. Making Highly Elastic and Tough Hydrogels
  from Doughs. \emph{Adv. Mater.} \textbf{2022}, \emph{34}, e2206577\relax
\mciteBstWouldAddEndPuncttrue
\mciteSetBstMidEndSepPunct{\mcitedefaultmidpunct}
{\mcitedefaultendpunct}{\mcitedefaultseppunct}\relax
\EndOfBibitem
\bibitem[Baumberger \latin{et~al.}(2006)Baumberger, Caroli, and
  Martina]{Martina:2006}
Baumberger,~T.; Caroli,~C.; Martina,~D. Solvent control of crack dynamics in a
  reversible hydrogel. \emph{Nat. Mater.} \textbf{2006}, \emph{5}, 552--5\relax
\mciteBstWouldAddEndPuncttrue
\mciteSetBstMidEndSepPunct{\mcitedefaultmidpunct}
{\mcitedefaultendpunct}{\mcitedefaultseppunct}\relax
\EndOfBibitem
\bibitem[Brown(2007)]{Brown:2007}
Brown,~H. A Model of the Fracture of Double Network Gels. \emph{Macromolecules}
  \textbf{2007}, \emph{40}, 3815–3818\relax
\mciteBstWouldAddEndPuncttrue
\mciteSetBstMidEndSepPunct{\mcitedefaultmidpunct}
{\mcitedefaultendpunct}{\mcitedefaultseppunct}\relax
\EndOfBibitem
\bibitem[Tanaka(2007)]{Tanaka:2007}
Tanaka,~Y. A local damage model for anomalous high toughness of double-network
  gels. \emph{Europhys. Lett.} \textbf{2007}, \emph{78}, 56005\relax
\mciteBstWouldAddEndPuncttrue
\mciteSetBstMidEndSepPunct{\mcitedefaultmidpunct}
{\mcitedefaultendpunct}{\mcitedefaultseppunct}\relax
\EndOfBibitem
\bibitem[Zhang \latin{et~al.}(2015)Zhang, Lin, Yuk, and Zhao]{Zhang:2015}
Zhang,~T.; Lin,~S.; Yuk,~H.; Zhao,~X. Predicting fracture energies and
  crack-tip fields of soft tough materials. \emph{Extreme Mech. Lett.}
  \textbf{2015}, \emph{4}, 1--8\relax
\mciteBstWouldAddEndPuncttrue
\mciteSetBstMidEndSepPunct{\mcitedefaultmidpunct}
{\mcitedefaultendpunct}{\mcitedefaultseppunct}\relax
\EndOfBibitem
\bibitem[Hui \latin{et~al.}(2003)Hui, Jagota, Bennison, and
  Londono]{Jagota:2003}
Hui,~C.-Y.; Jagota,~A.; Bennison,~S.; Londono,~J. Crack blunting and the
  strength of soft elastic solids. \emph{Proc. R. Soc. A} \textbf{2003},
  \emph{459}, 1489--1516\relax
\mciteBstWouldAddEndPuncttrue
\mciteSetBstMidEndSepPunct{\mcitedefaultmidpunct}
{\mcitedefaultendpunct}{\mcitedefaultseppunct}\relax
\EndOfBibitem
\bibitem[Seitz \latin{et~al.}(2009)Seitz, Martina, Baumberger, Krishnan, Hui,
  and Shull]{Seitz:2009}
Seitz,~M.~E.; Martina,~D.; Baumberger,~T.; Krishnan,~V.~R.; Hui,~C.-Y.;
  Shull,~K.~R. Fracture and large strain behavior of self-assembled triblock
  copolymer gels. \emph{Soft Matter} \textbf{2009}, \emph{5}, 447--456\relax
\mciteBstWouldAddEndPuncttrue
\mciteSetBstMidEndSepPunct{\mcitedefaultmidpunct}
{\mcitedefaultendpunct}{\mcitedefaultseppunct}\relax
\EndOfBibitem
\bibitem[Kim \latin{et~al.}(2020)Kim, Liu, Weon, Cohen, Hui, Dufresne, and
  Style]{Kimeaaz0418}
Kim,~J.~Y.; Liu,~Z.; Weon,~B.~M.; Cohen,~T.; Hui,~C.-Y.; Dufresne,~E.~R.;
  Style,~R.~W. Extreme cavity expansion in soft solids: Damage without
  fracture. \emph{Sci. Adv.} \textbf{2020}, \emph{6}, eaaz0418\relax
\mciteBstWouldAddEndPuncttrue
\mciteSetBstMidEndSepPunct{\mcitedefaultmidpunct}
{\mcitedefaultendpunct}{\mcitedefaultseppunct}\relax
\EndOfBibitem
\bibitem[Binnig \latin{et~al.}(1986)Binnig, Quate, and Gerber]{AFM}
Binnig,~G.; Quate,~C.~F.; Gerber,~C. Atomic Force Microscope. \emph{Phys. Rev.
  Lett.} \textbf{1986}, \emph{56}, 930--933\relax
\mciteBstWouldAddEndPuncttrue
\mciteSetBstMidEndSepPunct{\mcitedefaultmidpunct}
{\mcitedefaultendpunct}{\mcitedefaultseppunct}\relax
\EndOfBibitem
\bibitem[Xu \latin{et~al.}(2002)Xu, Zhang, and Zhang]{AFM_macromolecules_1}
Xu,~Q.; Zhang,~W.; Zhang,~X. Oxygen bridge inhibits conformational transition
  of 1,4-linked alpha-D-galactose detected by single-molecule atomic force
  microscopy. \emph{Macromolecules} \textbf{2002}, \emph{35}, 871--876\relax
\mciteBstWouldAddEndPuncttrue
\mciteSetBstMidEndSepPunct{\mcitedefaultmidpunct}
{\mcitedefaultendpunct}{\mcitedefaultseppunct}\relax
\EndOfBibitem
\bibitem[Gunari \latin{et~al.}(2006)Gunari, Schmidt, and
  Janshoff]{AFM_macromolecules_2}
Gunari,~N.; Schmidt,~M.; Janshoff,~A. Persistence length of cylindrical brush
  molecules measured by atomic force microscopy. \emph{Macromolecules}
  \textbf{2006}, \emph{39}, 2219--2224\relax
\mciteBstWouldAddEndPuncttrue
\mciteSetBstMidEndSepPunct{\mcitedefaultmidpunct}
{\mcitedefaultendpunct}{\mcitedefaultseppunct}\relax
\EndOfBibitem
\bibitem[Yang \latin{et~al.}(2018)Yang, Song, Feng, and
  Zhang]{AFM_macromolecules_3}
Yang,~P.; Song,~Y.; Feng,~W.; Zhang,~W. Unfolding of a Single Polymer Chain
  from the Single Crystal by Air-Phase Single-Molecule Force Spectroscopy:
  Toward Better Force Precision and More Accurate Description of Molecular
  Behaviors. \emph{Macromolecules} \textbf{2018}, \emph{51}, 7052--7060\relax
\mciteBstWouldAddEndPuncttrue
\mciteSetBstMidEndSepPunct{\mcitedefaultmidpunct}
{\mcitedefaultendpunct}{\mcitedefaultseppunct}\relax
\EndOfBibitem
\bibitem[Ashkin \latin{et~al.}(1986)Ashkin, Dziedzic, Bjorkholm, and
  Chu]{optical_tweezers}
Ashkin,~A.; Dziedzic,~J.~M.; Bjorkholm,~J.~E.; Chu,~S. Observation of a
  single-beam gradient force optical trap for dielectric particles. \emph{Opt.
  Lett.} \textbf{1986}, \emph{11}, 288--290\relax
\mciteBstWouldAddEndPuncttrue
\mciteSetBstMidEndSepPunct{\mcitedefaultmidpunct}
{\mcitedefaultendpunct}{\mcitedefaultseppunct}\relax
\EndOfBibitem
\bibitem[Wang \latin{et~al.}(1997)Wang, Yin, Landick, Gelles, and
  Block]{EWLC_marko_siggia}
Wang,~M.; Yin,~H.; Landick,~R.; Gelles,~J.; Block,~S. Stretching DNA with
  optical tweezers. \emph{Biophys. J.} \textbf{1997}, \emph{72},
  1335--1346\relax
\mciteBstWouldAddEndPuncttrue
\mciteSetBstMidEndSepPunct{\mcitedefaultmidpunct}
{\mcitedefaultendpunct}{\mcitedefaultseppunct}\relax
\EndOfBibitem
\bibitem[Rocha \latin{et~al.}(2018)Rocha, Storm, Bazoni, Ramos,
  Hernandez-Garcia, Stuart, Leermakers, and de~Vries]{AFM_OT_macromolecules}
Rocha,~M.~S.; Storm,~I.~M.; Bazoni,~R.~F.; Ramos,~E.~B.; Hernandez-Garcia,~A.;
  Stuart,~M. A.~C.; Leermakers,~F.; de~Vries,~R. Force and Scale Dependence of
  the Elasticity of Self-Assembled DNA Bottle Brushes. \emph{Macromolecules}
  \textbf{2018}, \emph{51}, 204--212\relax
\mciteBstWouldAddEndPuncttrue
\mciteSetBstMidEndSepPunct{\mcitedefaultmidpunct}
{\mcitedefaultendpunct}{\mcitedefaultseppunct}\relax
\EndOfBibitem
\bibitem[Smith \latin{et~al.}(1992)Smith, Finzi, and Bustamante]{FJC_smith}
Smith,~S.~B.; Finzi,~L.; Bustamante,~C. Direct Mechanical Measurements of the
  Elasticity of Single DNA Molecules by Using Magnetic Beads. \emph{Science}
  \textbf{1992}, \emph{258}, 1122--1126\relax
\mciteBstWouldAddEndPuncttrue
\mciteSetBstMidEndSepPunct{\mcitedefaultmidpunct}
{\mcitedefaultendpunct}{\mcitedefaultseppunct}\relax
\EndOfBibitem
\bibitem[Strick \latin{et~al.}(1996)Strick, Allemand, Bensimon, Bensimon, and
  Croquette]{magnetic_tweezers}
Strick,~T.; Allemand,~J.; Bensimon,~D.; Bensimon,~A.; Croquette,~V. The
  Elasticity of a Single Supercoiled DNA Molecule. \emph{Science}
  \textbf{1996}, \emph{271}, 1835--7\relax
\mciteBstWouldAddEndPuncttrue
\mciteSetBstMidEndSepPunct{\mcitedefaultmidpunct}
{\mcitedefaultendpunct}{\mcitedefaultseppunct}\relax
\EndOfBibitem
\bibitem[del Rio \latin{et~al.}(2009)del Rio, Perez-Jimenez, Liu, Roca-Cusachs,
  Fernandez, and Sheetz]{AFM_MT}
del Rio,~A.; Perez-Jimenez,~R.; Liu,~R.; Roca-Cusachs,~P.; Fernandez,~J.~M.;
  Sheetz,~M.~P. Stretching Single Talin Rod Molecules Activates Vinculin
  Binding. \emph{Science} \textbf{2009}, \emph{323}, 638--641\relax
\mciteBstWouldAddEndPuncttrue
\mciteSetBstMidEndSepPunct{\mcitedefaultmidpunct}
{\mcitedefaultendpunct}{\mcitedefaultseppunct}\relax
\EndOfBibitem
\bibitem[Butt \latin{et~al.}(2005)Butt, Cappella, and Kappl]{Butt_2005}
Butt,~H.-J.; Cappella,~B.; Kappl,~M. Force measurements with the atomic force
  microscope: Technique, interpretation and applications. \emph{Surf. Sci.
  Rep.} \textbf{2005}, \emph{59}, 1--152\relax
\mciteBstWouldAddEndPuncttrue
\mciteSetBstMidEndSepPunct{\mcitedefaultmidpunct}
{\mcitedefaultendpunct}{\mcitedefaultseppunct}\relax
\EndOfBibitem
\bibitem[Rief \latin{et~al.}(1997)Rief, Gautel, Fern, and
  Gaub]{Rief97reversibleunfolding}
Rief,~M.; Gautel,~M.; Fern,~J.~M.; Gaub,~H.~E. reversible unfolding of
  individual titin immunoglobulin dimains by AFM. \emph{Science} \textbf{1997},
  1109--1112\relax
\mciteBstWouldAddEndPuncttrue
\mciteSetBstMidEndSepPunct{\mcitedefaultmidpunct}
{\mcitedefaultendpunct}{\mcitedefaultseppunct}\relax
\EndOfBibitem
\bibitem[Fisher \latin{et~al.}(1999)Fisher, Oberhauser, Carrion-Vazquez,
  Marszalek, and Fernandez]{Fisher_2000}
Fisher,~T.; Oberhauser,~A.; Carrion-Vazquez,~M.; Marszalek,~P.; Fernandez,~J.
  The study of protein mechanics with the atomic force microscope. \emph{Trends
  Biochem. Sci.} \textbf{1999}, \emph{25}, 379--84\relax
\mciteBstWouldAddEndPuncttrue
\mciteSetBstMidEndSepPunct{\mcitedefaultmidpunct}
{\mcitedefaultendpunct}{\mcitedefaultseppunct}\relax
\EndOfBibitem
\bibitem[Rief \latin{et~al.}(1999)Rief, Clausen‐Schaumann, and
  Gaub]{Rief_1999}
Rief,~M.; Clausen‐Schaumann,~H.; Gaub,~H.~E. Sequence-dependent mechanics of
  single DNA molecules. \emph{Nat. Struct. Biol.} \textbf{1999}, \emph{6},
  346--349\relax
\mciteBstWouldAddEndPuncttrue
\mciteSetBstMidEndSepPunct{\mcitedefaultmidpunct}
{\mcitedefaultendpunct}{\mcitedefaultseppunct}\relax
\EndOfBibitem
\bibitem[Bustamante \latin{et~al.}(2000)Bustamante, Smith, Liphardt, and
  Smith]{Bustamante_2000}
Bustamante,~C.; Smith,~S.~B.; Liphardt,~J.; Smith,~D. Single-molecule studies
  of DNA mechanics. \emph{Curr. Opin. Struct. Biol.} \textbf{2000}, \emph{10},
  279--285\relax
\mciteBstWouldAddEndPuncttrue
\mciteSetBstMidEndSepPunct{\mcitedefaultmidpunct}
{\mcitedefaultendpunct}{\mcitedefaultseppunct}\relax
\EndOfBibitem
\bibitem[Rief \latin{et~al.}(1997)Rief, Oesterhelt, Heymann, and
  Gaub]{AFM_polysaccharides}
Rief,~M.; Oesterhelt,~F.; Heymann,~B.; Gaub,~H.~E. Single molecule force
  spectroscopy on polysaccharides by atomic force microscopy. \emph{Science}
  \textbf{1997}, \emph{275}, 1295--1297\relax
\mciteBstWouldAddEndPuncttrue
\mciteSetBstMidEndSepPunct{\mcitedefaultmidpunct}
{\mcitedefaultendpunct}{\mcitedefaultseppunct}\relax
\EndOfBibitem
\bibitem[Oesterhelt \latin{et~al.}(1999)Oesterhelt, Rief, and
  Gaub]{Oesterhelt_1999}
Oesterhelt,~F.; Rief,~M.; Gaub,~H.~E. Single molecule force spectroscopy by
  {AFM} indicates helical structure of poly(ethylene-glycol) in water.
  \emph{New J. Phys.} \textbf{1999}, \emph{1}, 6\relax
\mciteBstWouldAddEndPuncttrue
\mciteSetBstMidEndSepPunct{\mcitedefaultmidpunct}
{\mcitedefaultendpunct}{\mcitedefaultseppunct}\relax
\EndOfBibitem
\bibitem[Ortiz and Hadziioannou(1999)Ortiz, and
  Hadziioannou]{AFM_PMAA_macromolecules}
Ortiz,~C.; Hadziioannou,~G. Entropic elasticity of single polymer chains of
  poly(methacrylic acid) measured by atomic force microscopy.
  \emph{Macromolecules} \textbf{1999}, \emph{32}, 780--787\relax
\mciteBstWouldAddEndPuncttrue
\mciteSetBstMidEndSepPunct{\mcitedefaultmidpunct}
{\mcitedefaultendpunct}{\mcitedefaultseppunct}\relax
\EndOfBibitem
\bibitem[Schwaderer \latin{et~al.}(2008)Schwaderer, Funk, Achenbach, Weis,
  Bräuchle, and Michaelis]{AFM_PDMS}
Schwaderer,~P.; Funk,~E.; Achenbach,~F.; Weis,~J.; Bräuchle,~C.; Michaelis,~J.
  Single-Molecule Measurement of the Strength of a Siloxane Bond †.
  \emph{Langmuir} \textbf{2008}, \emph{24}, 1343--9\relax
\mciteBstWouldAddEndPuncttrue
\mciteSetBstMidEndSepPunct{\mcitedefaultmidpunct}
{\mcitedefaultendpunct}{\mcitedefaultseppunct}\relax
\EndOfBibitem
\bibitem[Kresse and Hafner(1993)Kresse, and Hafner]{AIMD}
Kresse,~G.; Hafner,~J. Ab initio molecular dynamics for liquid metals.
  \emph{Phys. Rev. B} \textbf{1993}, \emph{47}, 558--561\relax
\mciteBstWouldAddEndPuncttrue
\mciteSetBstMidEndSepPunct{\mcitedefaultmidpunct}
{\mcitedefaultendpunct}{\mcitedefaultseppunct}\relax
\EndOfBibitem
\bibitem[Sun(1995)]{AIMD_macromolecules}
Sun,~H. Ab initio calculations and force field development for computer
  simulation of polysilanes. \emph{Macromolecules} \textbf{1995}, \emph{28},
  701--712\relax
\mciteBstWouldAddEndPuncttrue
\mciteSetBstMidEndSepPunct{\mcitedefaultmidpunct}
{\mcitedefaultendpunct}{\mcitedefaultseppunct}\relax
\EndOfBibitem
\bibitem[Lupton \latin{et~al.}(2005)Lupton, Nonnenberg, Frank, Achenbach, Weis,
  and Bräuchle]{AIMD_PDMS_single}
Lupton,~E.; Nonnenberg,~C.; Frank,~I.; Achenbach,~F.; Weis,~J.; Bräuchle,~C.
  Stretching siloxanes: An ab initio molecular dynamics study. \emph{Chem.
  Phys. Lett.} \textbf{2005}, \emph{414}, 132--137\relax
\mciteBstWouldAddEndPuncttrue
\mciteSetBstMidEndSepPunct{\mcitedefaultmidpunct}
{\mcitedefaultendpunct}{\mcitedefaultseppunct}\relax
\EndOfBibitem
\bibitem[Lupton \latin{et~al.}(2009)Lupton, Achenbach, Weis, Bräuchle, and
  Frank]{AIMD_PDMS_multiple}
Lupton,~E.; Achenbach,~F.; Weis,~J.; Bräuchle,~C.; Frank,~I. Origins of
  Material Failure in Siloxane Elastomers from First Principles.
  \emph{Chem{P}hys{C}hem} \textbf{2009}, \emph{10}, 119--23\relax
\mciteBstWouldAddEndPuncttrue
\mciteSetBstMidEndSepPunct{\mcitedefaultmidpunct}
{\mcitedefaultendpunct}{\mcitedefaultseppunct}\relax
\EndOfBibitem
\bibitem[Lupton \latin{et~al.}(2006)Lupton, Achenbach, Weis, Bräuchle, and
  Frank]{AIMD_PDMS_HMDSO}
Lupton,~E.; Achenbach,~F.; Weis,~J.; Bräuchle,~C.; Frank,~I. Modified
  Chemistry of Siloxanes under Tensile Stress: Interaction with Environment.
  \emph{J. Phys. Chem. B} \textbf{2006}, \emph{110}, 14557--63\relax
\mciteBstWouldAddEndPuncttrue
\mciteSetBstMidEndSepPunct{\mcitedefaultmidpunct}
{\mcitedefaultendpunct}{\mcitedefaultseppunct}\relax
\EndOfBibitem
\bibitem[Lu \latin{et~al.}(1998)Lu, Isralewitz, Krammer, Vogel, and
  Schulten]{LU1998}
Lu,~H.; Isralewitz,~B.; Krammer,~A.; Vogel,~V.; Schulten,~K. Unfolding of Titin
  Immunoglobulin Domains by Steered Molecular Dynamics Simulation.
  \emph{Biophys. J.} \textbf{1998}, \emph{75}, 662--671\relax
\mciteBstWouldAddEndPuncttrue
\mciteSetBstMidEndSepPunct{\mcitedefaultmidpunct}
{\mcitedefaultendpunct}{\mcitedefaultseppunct}\relax
\EndOfBibitem
\bibitem[Lu and Schulten(1999)Lu, and Schulten]{LU1999}
Lu,~H.; Schulten,~K. Steered molecular dynamics simulations of force-induced
  protein domain unfolding. \emph{Proteins: Struct. Function Bioinf.}
  \textbf{1999}, \emph{35}, 453--463\relax
\mciteBstWouldAddEndPuncttrue
\mciteSetBstMidEndSepPunct{\mcitedefaultmidpunct}
{\mcitedefaultendpunct}{\mcitedefaultseppunct}\relax
\EndOfBibitem
\bibitem[Isralewitz \latin{et~al.}(2001)Isralewitz, Baudry, Gullingsrud,
  Kosztin, and Schulten]{ISRALEWITZ2001}
Isralewitz,~B.; Baudry,~J.; Gullingsrud,~J.; Kosztin,~D.; Schulten,~K. Steered
  molecular dynamics investigations of protein function. \emph{J. Molec. Graph.
  Model.} \textbf{2001}, \emph{19}, 13--25\relax
\mciteBstWouldAddEndPuncttrue
\mciteSetBstMidEndSepPunct{\mcitedefaultmidpunct}
{\mcitedefaultendpunct}{\mcitedefaultseppunct}\relax
\EndOfBibitem
\bibitem[Han \latin{et~al.}(2022)Han, Hilburg, and
  Alexander-Katz]{SMD_macromolecules}
Han,~Z.; Hilburg,~S.~L.; Alexander-Katz,~A. Forced Unfolding of
  Protein-Inspired Single-Chain Random Heteropolymers. \emph{Macromolecules}
  \textbf{2022}, \emph{55}, 1295--1309\relax
\mciteBstWouldAddEndPuncttrue
\mciteSetBstMidEndSepPunct{\mcitedefaultmidpunct}
{\mcitedefaultendpunct}{\mcitedefaultseppunct}\relax
\EndOfBibitem
\bibitem[{Van Duin} \latin{et~al.}(2001){Van Duin}, Dasgupta, Lorant, and
  Goddard]{ReaxFF}
{Van Duin},~A.; Dasgupta,~S.; Lorant,~F.; Goddard,~W. ReaxFF: A reactive force
  field for hydrocarbons. \emph{J. Phys. Chem. A} \textbf{2001}, \emph{105},
  9396--9409\relax
\mciteBstWouldAddEndPuncttrue
\mciteSetBstMidEndSepPunct{\mcitedefaultmidpunct}
{\mcitedefaultendpunct}{\mcitedefaultseppunct}\relax
\EndOfBibitem
\bibitem[Chenoweth \latin{et~al.}(2005)Chenoweth, Cheung, van Duin, Goddard,
  and Kober]{PDMS_ReaxFF}
Chenoweth,~K.; Cheung,~S.; van Duin,~A.; Goddard,~W.; Kober,~E. Simulations on
  the Thermal Decomposition of a Poly(dimethylsiloxane) Polymer Using the
  ReaxFF Reactive Force Field. \emph{J. Amer. Chem. Soc.} \textbf{2005},
  \emph{127}, 7192--202\relax
\mciteBstWouldAddEndPuncttrue
\mciteSetBstMidEndSepPunct{\mcitedefaultmidpunct}
{\mcitedefaultendpunct}{\mcitedefaultseppunct}\relax
\EndOfBibitem
\bibitem[Newsome \latin{et~al.}(2012)Newsome, Sengupta, Foroutan, Russo, and
  Duin]{Newsome2012}
Newsome,~D.~A.; Sengupta,~D.; Foroutan,~H.; Russo,~M.~F.; Duin,~A. C.~T.
  Oxidation of Silicon Carbide by O2 and H2O: A ReaxFF Reactive Molecular
  Dynamics Study, Part I. \emph{J. Phys. Chem. C} \textbf{2012}, \emph{116},
  16111--16121\relax
\mciteBstWouldAddEndPuncttrue
\mciteSetBstMidEndSepPunct{\mcitedefaultmidpunct}
{\mcitedefaultendpunct}{\mcitedefaultseppunct}\relax
\EndOfBibitem
\bibitem[Soria \latin{et~al.}(2017)Soria, Zhang, van Duin, and
  Patrito]{Soria2017}
Soria,~F.~A.; Zhang,~W.; van Duin,~A. C.~T.; Patrito,~E.~M. Thermal Stability
  of Organic Monolayers Grafted to Si(111): Insights from ReaxFF Reactive
  Molecular Dynamics Simulations. \emph{ACS Appl. Mater. Interf.}
  \textbf{2017}, \emph{9}, 30969--30981\relax
\mciteBstWouldAddEndPuncttrue
\mciteSetBstMidEndSepPunct{\mcitedefaultmidpunct}
{\mcitedefaultendpunct}{\mcitedefaultseppunct}\relax
\EndOfBibitem
\bibitem[Soria \latin{et~al.}(2018-10-18)Soria, Zhang, Paredes-Olivera, van
  Duin, and Patrito]{Soria2018}
Soria,~F.~A.; Zhang,~W.; Paredes-Olivera,~P.~A.; van Duin,~A. C.~T.;
  Patrito,~E.~M. Si/C/H ReaxFF Reactive Potential for Silicon Surfaces Grafted
  with Organic Molecules. \emph{J. Phys. Chem.} \textbf{2018-10-18},
  \emph{122}\relax
\mciteBstWouldAddEndPuncttrue
\mciteSetBstMidEndSepPunct{\mcitedefaultmidpunct}
{\mcitedefaultendpunct}{\mcitedefaultseppunct}\relax
\EndOfBibitem
\bibitem[Plimpton(1995)]{PLIMPTON19951}
Plimpton,~S. Fast Parallel Algorithms for Short-Range Molecular Dynamics.
  \emph{J. Comput. Phys.} \textbf{1995}, \emph{117}, 1--19\relax
\mciteBstWouldAddEndPuncttrue
\mciteSetBstMidEndSepPunct{\mcitedefaultmidpunct}
{\mcitedefaultendpunct}{\mcitedefaultseppunct}\relax
\EndOfBibitem
\bibitem[Smith \latin{et~al.}(1996)Smith, Cui, and Bustamante]{EFJC_smith}
Smith,~S.~B.; Cui,~Y.; Bustamante,~C. Overstretching B-DNA: The Elastic
  Response of Individual Double-Stranded and Single-Stranded DNA Molecules.
  \emph{Science} \textbf{1996}, \emph{271}, 795--799\relax
\mciteBstWouldAddEndPuncttrue
\mciteSetBstMidEndSepPunct{\mcitedefaultmidpunct}
{\mcitedefaultendpunct}{\mcitedefaultseppunct}\relax
\EndOfBibitem
\bibitem[Preston \latin{et~al.}(2021)Preston, Gelin, and Kosov]{Preston2021}
Preston,~R.~J.; Gelin,~M.~F.; Kosov,~D.~S. First-passage time theory of
  activated rate chemical processes in electronic molecular junctions. \emph{J.
  Chem. Phys.} \textbf{2021}, \emph{154}, 114108\relax
\mciteBstWouldAddEndPuncttrue
\mciteSetBstMidEndSepPunct{\mcitedefaultmidpunct}
{\mcitedefaultendpunct}{\mcitedefaultseppunct}\relax
\EndOfBibitem
\bibitem[Razbin \latin{et~al.}(2019)Razbin, Benetatos, and
  Moosavi-Movahedi]{Razbin2019}
Razbin,~M.; Benetatos,~P.; Moosavi-Movahedi,~A.~A. A first-passage approach to
  the thermal breakage of a discrete one-dimensional chain. \emph{Soft Matter}
  \textbf{2019}, \emph{15}, 2469--2478\relax
\mciteBstWouldAddEndPuncttrue
\mciteSetBstMidEndSepPunct{\mcitedefaultmidpunct}
{\mcitedefaultendpunct}{\mcitedefaultseppunct}\relax
\EndOfBibitem
\bibitem[Puthur and Sebastian(2002)Puthur, and Sebastian]{Puthur2002}
Puthur,~R.; Sebastian,~K. Theory of polymer breaking under tension. \emph{Phys.
  Rev. B} \textbf{2002}, \emph{66}, 024304\relax
\mciteBstWouldAddEndPuncttrue
\mciteSetBstMidEndSepPunct{\mcitedefaultmidpunct}
{\mcitedefaultendpunct}{\mcitedefaultseppunct}\relax
\EndOfBibitem
\bibitem[Berezhkovskii and Makarov(2019)Berezhkovskii, and
  Makarov]{Berezhkovskii2019}
Berezhkovskii,~A.~M.; Makarov,~D.~E. On distributions of barrier crossing times
  as observed in single-molecule studies of biomolecules. \emph{Biophys. Rep.}
  \textbf{2019}, \emph{1}, 100029\relax
\mciteBstWouldAddEndPuncttrue
\mciteSetBstMidEndSepPunct{\mcitedefaultmidpunct}
{\mcitedefaultendpunct}{\mcitedefaultseppunct}\relax
\EndOfBibitem
\bibitem[Risken(1996)]{Risken}
Risken,~H. \emph{The Fokker-Planck Equation}; Springer: Berlin, 1996\relax
\mciteBstWouldAddEndPuncttrue
\mciteSetBstMidEndSepPunct{\mcitedefaultmidpunct}
{\mcitedefaultendpunct}{\mcitedefaultseppunct}\relax
\EndOfBibitem
\bibitem[Kampen(2007)]{Kampen}
Kampen,~N.~V. \emph{Stochastic Processes in Physics and Chemistry, 3rd Ed.};
  North Holland: Amsterdam, The Netherlands, 2007\relax
\mciteBstWouldAddEndPuncttrue
\mciteSetBstMidEndSepPunct{\mcitedefaultmidpunct}
{\mcitedefaultendpunct}{\mcitedefaultseppunct}\relax
\EndOfBibitem
\bibitem[Gardiner(1985)]{Gardiner}
Gardiner,~C.~W. \emph{Handbook of Stochastic Methods, 2nd Ed.}; Springer,
  Berlin: Berlin, 1985\relax
\mciteBstWouldAddEndPuncttrue
\mciteSetBstMidEndSepPunct{\mcitedefaultmidpunct}
{\mcitedefaultendpunct}{\mcitedefaultseppunct}\relax
\EndOfBibitem
\bibitem[Dembo \latin{et~al.}(1988)Dembo, Torney, Saxman, and
  Hammer]{Danbo:1988}
Dembo,~M.; Torney,~D.; Saxman,~K.; Hammer,~D. The reaction-limited kinetics of
  membrane-to-surface adhesion and detachment. \emph{Proc. R. Soc. B}
  \textbf{1988}, \emph{234}, 55--83\relax
\mciteBstWouldAddEndPuncttrue
\mciteSetBstMidEndSepPunct{\mcitedefaultmidpunct}
{\mcitedefaultendpunct}{\mcitedefaultseppunct}\relax
\EndOfBibitem
\bibitem[Guo and Guilford(2006)Guo, and Guilford]{Guo:2006}
Guo,~B.; Guilford,~W. Mechanics of actomyosin bonds in different nucleotide
  states are tuned to muscle contraction. \emph{Proc. Natl. Acadm. Sci. USA}
  \textbf{2006}, \emph{103}, 9844--9\relax
\mciteBstWouldAddEndPuncttrue
\mciteSetBstMidEndSepPunct{\mcitedefaultmidpunct}
{\mcitedefaultendpunct}{\mcitedefaultseppunct}\relax
\EndOfBibitem
\bibitem[Liu \latin{et~al.}(2014)Liu, Chen, Evavold, and Zhu]{Liu:2014}
Liu,~B.; Chen,~W.; Evavold,~B.; Zhu,~C. Accumulation of dynamic catch bonds
  between TCR and agonist peptide-MHC triggers T cell signaling. \emph{Cell}
  \textbf{2014}, \emph{157}, 357--68\relax
\mciteBstWouldAddEndPuncttrue
\mciteSetBstMidEndSepPunct{\mcitedefaultmidpunct}
{\mcitedefaultendpunct}{\mcitedefaultseppunct}\relax
\EndOfBibitem
\bibitem[Das \latin{et~al.}(2016)Das, Mallis, Duke-Cohan, Hussey, Tetteh,
  Hilton, Wagner, Lang, and Reinherz]{Reinherz:2016}
Das,~D.; Mallis,~R.; Duke-Cohan,~J.; Hussey,~R.; Tetteh,~P.; Hilton,~M.;
  Wagner,~G.; Lang,~M.; Reinherz,~E. Pre-T Cell Receptors (Pre-TCRs) Leverage
  Vbeta Complementarity Determining Regions (CDRs) and Hydrophobic Patch in
  Mechanosensing Thymic Self-ligands. \emph{J. Biolog. Chem.} \textbf{2016},
  \emph{291}, 25292–305\relax
\mciteBstWouldAddEndPuncttrue
\mciteSetBstMidEndSepPunct{\mcitedefaultmidpunct}
{\mcitedefaultendpunct}{\mcitedefaultseppunct}\relax
\EndOfBibitem
\bibitem[Flowers and Switzer(1978)Flowers, and Switzer]{Sylgard-184/186}
Flowers,~G.~L.; Switzer,~S.~T. Background material properties of selected
  silicone potting compounds and raw materials for their substitutes.
  \textbf{1978}, United States Government document\relax
\mciteBstWouldAddEndPuncttrue
\mciteSetBstMidEndSepPunct{\mcitedefaultmidpunct}
{\mcitedefaultendpunct}{\mcitedefaultseppunct}\relax
\EndOfBibitem
\bibitem[Evmenenko \latin{et~al.}(2006)Evmenenko, Mo, Kewalramani, and
  Dutta]{Evmenenko:2005}
Evmenenko,~G.; Mo,~H.; Kewalramani,~S.; Dutta,~P. Conformational rearrangements
  in interfacial region of polydimethylsiloxane melt films. \emph{Polymer}
  \textbf{2006}, \emph{47}, 878--882\relax
\mciteBstWouldAddEndPuncttrue
\mciteSetBstMidEndSepPunct{\mcitedefaultmidpunct}
{\mcitedefaultendpunct}{\mcitedefaultseppunct}\relax
\EndOfBibitem
\end{mcitethebibliography}

\clearpage

\section*{\sc Supplementary Information}

\parindent 0mm

{\Large \bf Siloxane molecules: Nonlinear elastic behavior and fracture characteristics}

Tianchi Li$^{1}$, Eric R. Dufresne$^{1}$, Martin Kr\"oger$^{2,3}$, Stefanie Heyden$^{1*)}$\\[5mm]
$^{1)}$ Soft and Living Materials, Department of Materials, ETH Zurich, CH--8093 Zurich, Switzerland

$^{2)}$ Polymer Physics, Department of Materials, ETH Zurich, CH--8093 Zurich, Switzerland

$^{3)}$ Magnetism and Interface Physics, Department of Materials, ETH Zurich, CH--8093 Zurich, Switzerland
\\[5mm]
$^{*)}$ Corresponding author: stefanie.heyden@mat.ethz.ch (S.H.)

\setcounter{figure}{0}
\setcounter{equation}{0}
\setcounter{page}{1}
\setcounter{section}{0}
\setcounter{subsection}{0}
\renewcommand{\thefigure}{S\arabic{figure}}
\renewcommand{\thetable}{S\Roman{table}}
\renewcommand{\theequation}{S-\arabic{equation}}
\renewcommand{\thesubsection}{S\arabic{subsection}}

\newpage

\subsection{Characteristic force at finite temperature}

\begin{figure*}
    \centering
    \includegraphics[width=0.45\textwidth]{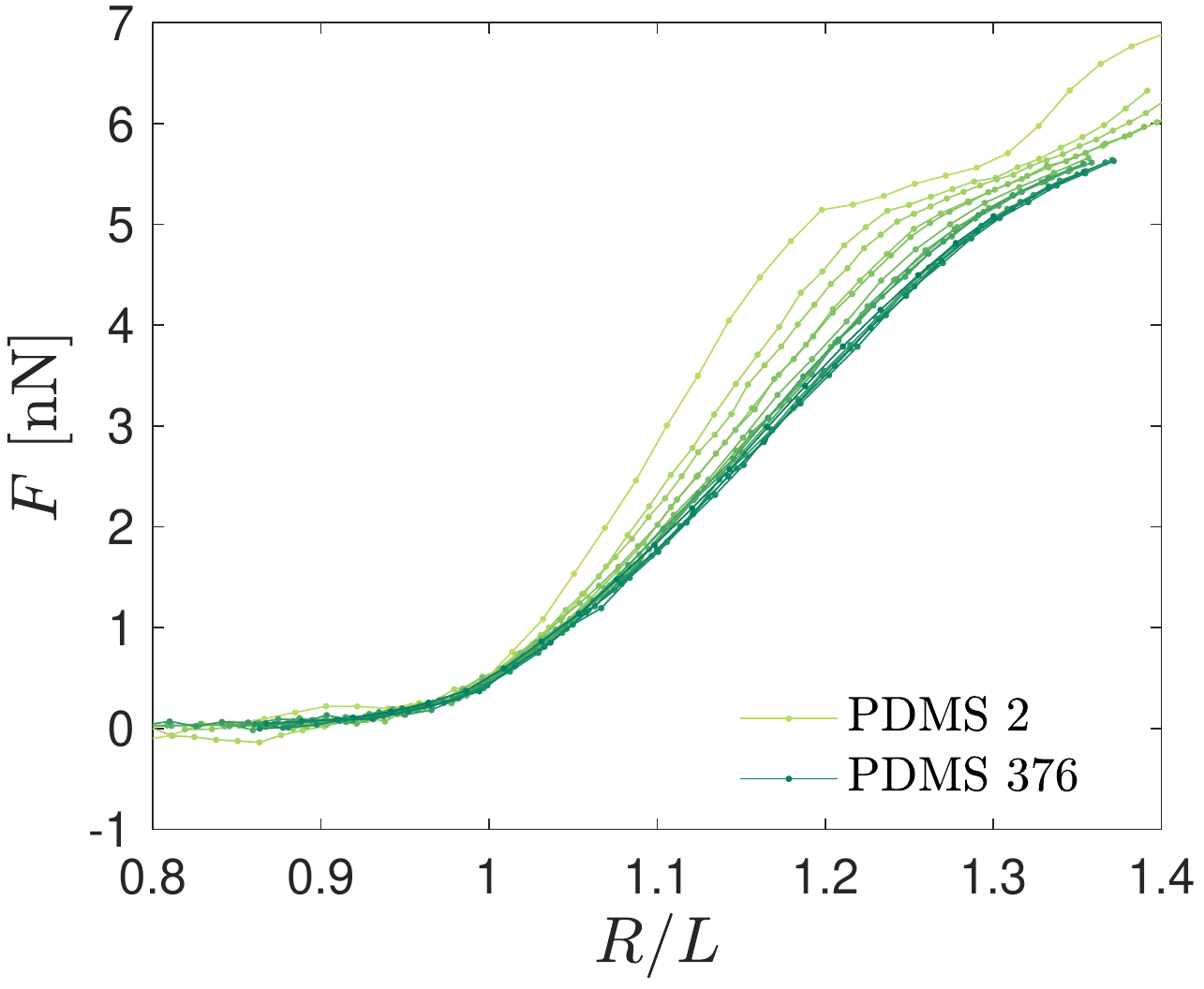}
    \includegraphics[width=0.462\textwidth]{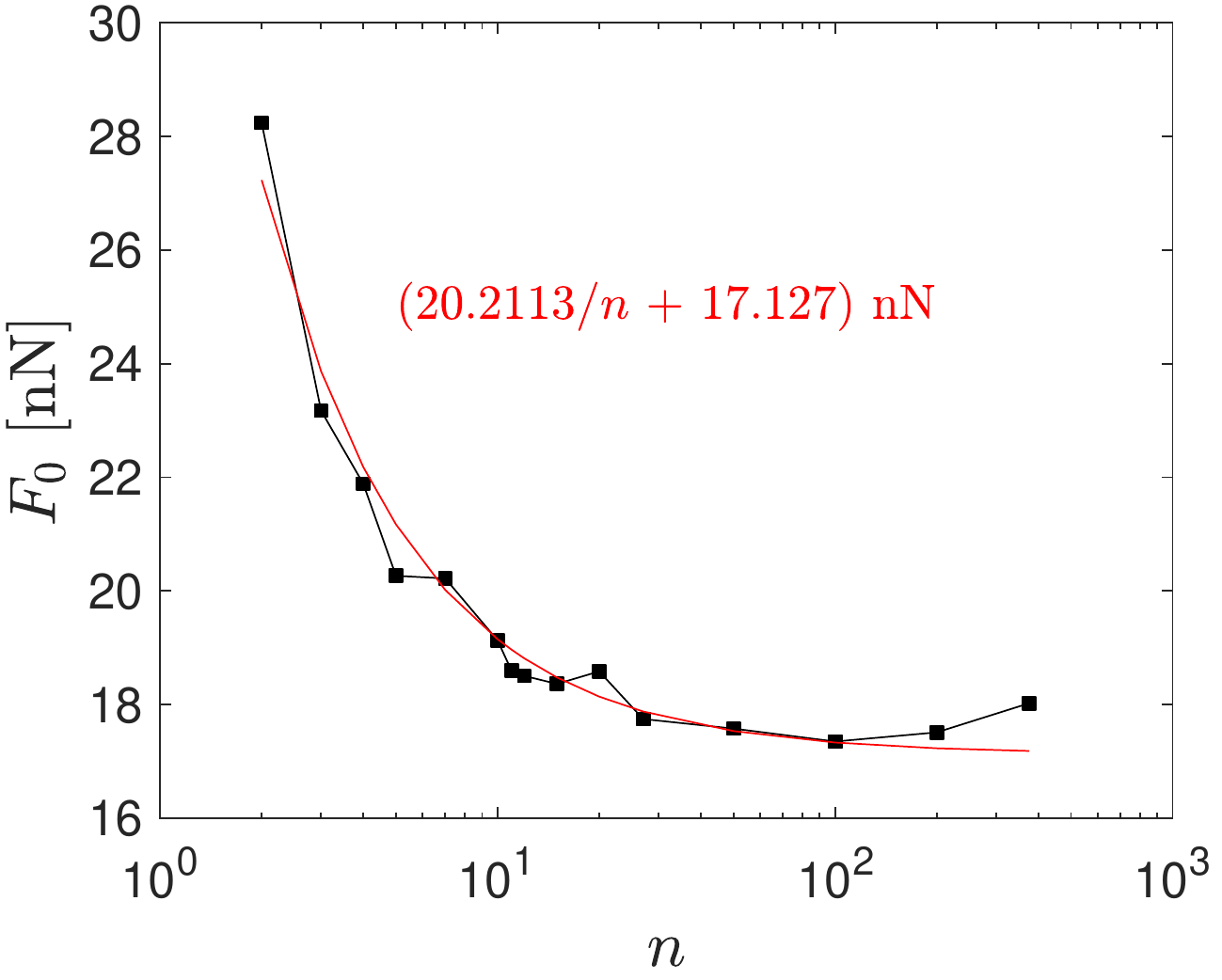}
    \caption{Characteristic force (equivalent to Fig. 1(d) in the main manuscript) at $T=300$\,K.
    The end effect (decreasing $F_0$ with increasing $n$) is also observed at finite temperature. 
    Added noise due to entropic contributions impede  the characterization of $F_0(n)$ within the enthalpic region, for which we still recover $F_0\propto (n+1)/n$. 
    In combination with a study of bond length distributions along the chain (for all forces in the enthalpic part, the two terminal Kuhn springs (Si-Si-Si) are longer than all internal springs), this leads us to the non-uniform chain of springs model.}
    \label{fig-s1}
\end{figure*}

\subsection{Kuhn length}
\label{appendix-Kuhn}

\begin{figure*}
    \centering
    \includegraphics[width=0.54\textwidth]{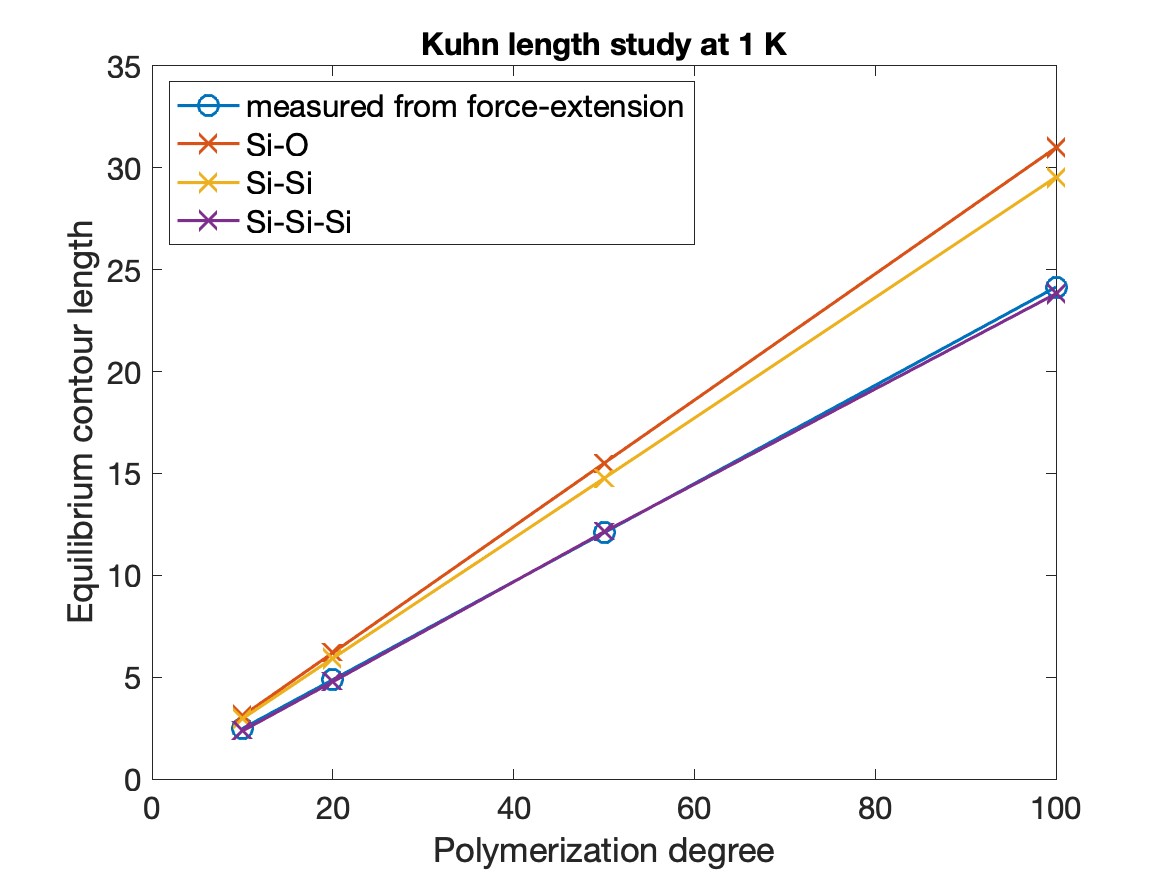}
    \caption{Measured equilibrium contour length (blue circles) in comparison to different models of Kuhn segments (Si-O, Si-Si, and Si-Si-Si).
    Using force-extension simulations at $T=1$\,K, the measured equilibrium contour length (specified in units of nm) follows from the intercept of the enthalpic part on the horizontal axis at $F=0$.
    Following the EFJC model, the equilibrium contour length is the sum of all Kuhn segment lengths.
    We find that taking Si-O or Si-Si as a Kuhn segment overestimates the measured equilibrium contour length, while taking Si-Si-Si as the elementary Kuhn unit is in good agreement with measured values.}
    \label{fig:Kuhn_length}
\end{figure*}

We obtained an independent estimate of the Kuhn length of the PDMS chains by analyzing the Si-Si vector correlation function $C(i)=\langle{\bf u}_j\cdot{\bf u}_{j+i}\rangle$, 
where ${\bf u}_j$ denotes the unit vector parallel to the vector connecting the $j$th and $(j+1)$th Si atom along the PDMS backbone, and the average is taken over all $j\in\{1,..,n\}$ within an ensemble of equilibrium PDMS-$n$ chains. For the FRC model, $\ln C(i)=-\ell/L_p$, where $L_p$ is the persistence length, and $\ell\approx 2.932\pm 0.003$\,\AA{} the measured average distance between adjacent Si atoms. We obtain $L_p=2.67\pm 0.07$\,\AA{} ($n=10$), $L_p=2.86\pm 0.09$\,\AA{} ($n=20$), $L_p=2.69\pm 0.06$\,\AA{} ($n=50$), $L_p=2.77\pm 0.05$\,\AA{} ($n=100$). The Kuhn length $L_k$ is twice as large as the persistence length, confirming $L_k\approx 5.5$\,\AA. This is in agreement with prior literature estimates \cite{Evmenenko:2005}.

\newpage

\subsection{Parameters of the double well potential}
\begin{table*}[h!]
\scriptsize
    \begin{tabular}{lrcccl}
    \hline\hline 
                & $b_r$ & $k_2$ & $c_2$ & $b_2$ & $b_1$   \\
      System    & [nm] & [kg.nm$^{-2}$.s$^{-2}$] & [kg.s$^{-2}$] & [nm] & [nm]  $\qquad\tilde{F}\equiv F/$ nN \\
     \hline 
      PDMS 4 (8 Si-O bonds) & 0.22 & $579866$ & $-55.9156$ & $0.186$ & $0.15471 + 0.00085126\,\tilde{F} + 0.0002648\,\tilde{F}^2$\\
      PDMS 376 (752 Si-O bonds) & 0.22 & $621285$ & $-41.419$ & $0.185$ & $0.15471 + 0.00085126\,\tilde{F} + 0.0002648\,\tilde{F}^2$\\
      linked PDMS 4 (2 Si-C bonds) & 0.255 & $0$ & $186$ & $0$ & $0.20042 + 0.0029917\,\tilde{F} + 0.00031558\,\tilde{F}^2$\\
      linked PDMS 4 (1 C-C bond) & $^{(*)}$0.181 & $0$ & $800$ & $0$ & $0.151 - 0.0005\,\tilde{F}$\\
      \hline\hline
    \end{tabular}
    \caption{Parameters of the effective bond potential $U(b;F)$ given by Eq.\ \eqref{eq:potential}. 
    All results have been produced using the stated values, whose error is less than 3\% for $k_2$ and $c_2$, and less than 1\% for $b_1$ and $b_2$. 
    For PDMS 4 (4 Si-O bonds) we obtained $\tau_4(F=90$ kcal.mol$^{-1}$.\AA$^{-1}=6.2531$ nN$)=0.221\pm 0.002$ ns from atomistic simulation at $T=300$ K.
    This value is reproduced via Brownian dynamics using $\zeta=105$ ng/m$^2$. For linked PDMS 4 (2 Si-C bonds) $b_r=0.255$ nm and we obtained $\tau_2(F=90$ kcal.mol$^{-1}$.\AA$^{-1})=0.70\pm 0.04$ ps and $\tau_1(F=80$ kcal.mol$^{-1}$.\AA$^{-1})=0.94\pm 0.05$ ps and $\tau_1(F=75$ kcal.mol$^{-1}$.\AA$^{-1})=51.3\pm 0.8$ ps from atomistic simulation at $T=300$ K. This value is reproduced via Brownian dynamics using $\zeta=0.002$ ng/m$^2$. $^{(*)}$ literature estimate.
    }
    \label{tab_parameters}
\end{table*}

\newpage

\subsection{Mean bond rupture time}

Consider a Brownian bond whose time-dependent length $b(t)$ resides within the interval $[b_0,b_r]$.
The bond is assumed to change its length resulting from three types of forces: Deterministic
forces due to a one-dimensional potential $U(b;F)$ of mean force, which we determine
from atomistic simulation at constant force $F$, a frictional force (friction coefficient $\zeta$) resulting from 
the surrounding medium, and a stochastic force whose strength is governed by the fluctuation-dissipation
theorem. The Langevin equation for the bond length $b$ thus reads \cite{Risken}
\be 
 \frac{d}{dt} b = -\frac{1}{\zeta} \frac{dU(b;F)}{db} + \sqrt{\frac{2\kB T}{\zeta}} \eta(t),
 \label{Langevinbond}
\ee 
where $\eta(t)$ represents uncorrelated white noise, $\langle\eta(t)\rangle=0$ and $\langle\eta(t)\eta(t')\rangle=\delta(t-t')$. 
We further consider
an adsorbing boundary at $b=b_r$ (the rupture bond length) and a reflecting boundary at $b=b_0$. Let 
the conditional probability $p_2(b',t|b,0)$ distribution capture the probability that 
a bond, whose length is $b$ at time $0$, assumes length $b'$ at a later time $t'\ge 0$, 
with $b,b'\in[b_0,b_r]$. Inline with our assumptions, 
$p_2(b',t|b,0)$ solves an adjungated Fokker-Planck equation corresponding to the Langevin Eq.\ \eqref{Langevinbond} \cite{Gardiner,Kampen}
\bea 
 -\frac{\partial}{\partial t} p_2(b',t|b,0) 
 &=& -\left[-\frac{1}{\zeta}\frac{\partial U(b;F)}{\partial  b}\frac{\partial}{\partial b} + \frac{\kB T}{\zeta} \frac{\partial^2}{\partial b^2}\right] p_2(b',t|b,0) \label{FPbond}
\eea 
subject to initial condition 
$p_2(b',0|b,0)=\delta(b'-b)$ and the abovementioned constraints. Then 
\be 
 G(b_r,t|b) = \int_{b_0}^{b_r} p_2(b',t|b,0) \,db' 
\ee 
is the probability that $b(t)$ resides within the interval $[b_0,b_r]$ at time $t$. 
The $G$ is thus not normalized except in the limit $b_r\rightarrow\infty$, i.e.,  
one has $G(\infty,t|b)=1$.  Further $G(b_r,0|b)=1$ since $p_2(b',0|b,0)=\delta(b'-b)$,
and $G(b_r,\infty|b)=0$, since the bond length exceeds $b_r$ with a nonzero probability.
One can write down an equation for $G$ based on the equation \eqref{FPbond} for $p_2$ \cite{Risken}. 
Due to the boundary conditions for $p_2$, the boundary conditions for $G$ read $G(b_r,0|b)=1$ 
for $b\in[b_0,b_r]$ and $G(b_r,0|b)=0$ otherwise, and
\be 
 \left.G(b_r,t|b)\right|_{b=b_r} = 0 , \qquad \left.\frac{\partial}{\partial b} G(b_r,t|b)\right|_{b=b_0} = 0.
\ee 

Because we are interested in the mean rupture time, we introduce the fraction $f(b_r,t|b)$ of bonds 
that reach $b_r$ (and thus leave the interval $[b_0,b_r]$) within the time interval $[t,t+dt]$. One has
\be 
 -dG(b_r,t|b) = -\partial_t G(b_r,t|b)dt \equiv f(b_r,t|b) dt, 
 \qquad f(b_r,t|b) = -\partial_t G(b_r,t|b) .
\ee 
The quantity 
\be 
 T_1(b_r,b) = \int_0^\infty t f(b_r,t|b) dt = -\int_0^\infty t \partial_t G(b_r,t|b) dt 
 = \int_0^\infty G(b_r,t|b) dt 
\ee 
is the mean bond rupture time. Higher moments can also be calculated with
the cumulative distribution function $G(b_r,t|b)$ at hand. 
The equation for $G$ can now be used to write down coupled equations for the moments
\be 
 T_j(b_r,b) = \int_0^\infty t^n f(b_r,t|b) = j \int_0^\infty t^{j-1} G(b_r,t|b) \qquad (j\ge 1, T_0=1).
\ee 
The equation for the $j$th moment reads
\be 
 \left[ \frac{\kB T}{\zeta} \frac{\partial^2}{\partial b^2} - \frac{1}{\zeta}\frac{\partial U(b;F)}{\partial b} \frac{\partial}{\partial b} \right] T_j(b_r,b) = -j T_{j-1}(b_r,b) \qquad j=1,2,..  \label{coupledT}
\ee 
and the boundary conditions for $T_j(b_r,b)$ are 
\be 
 T_j(b_r,b_r) = 0, \qquad \left.\frac{\partial}{\partial b} T_j(b_r,b)\right|_{b=b_0} = 0 .
\ee 
For $j=1$ the above Eq.\ \eqref{coupledT} reduces to 
\be 
\left[ \frac{\kB T}{\zeta} \frac{\partial^2}{\partial b^2} -\frac{1}{\zeta}\frac{\partial U(b;F)}{\partial b} \frac{\partial}{\partial b} \right] T_1(b_r,b) = -1 .
\ee 
This ordinary differential boundary problem is solved by 
\be 
 T_1(b_r,b) = \int_b^{b_r} dz\, \frac{1}{\Psi(z)} \int_{b_0}^z \frac{\Psi(y)}{D}\, dy,
 \label{finalT1}
\ee 
with
\be 
   \Psi(z) = 
   \exp\left[\int^z \frac{-\frac{1}{\zeta}\frac{\partial U(b;F)}{\partial b}}{\kB T/\zeta} db \right] 
    = \exp\left[-\frac{U(z;F)}{\kB T}\right].
 \label{finalPsi}
\ee 
If we average the mean rupture time over all possible initial lengths 
of the bond, 
\be 
 \overline{T_1}(b_r) = \frac{\int_{b_0}^{b_r} T_1(b_r,b) p_0(b) \, db}{\int_{b_0}^{b_r} p_0(b)\,db},
\ee
where $p_0(b)$ is the density distribution of the initial value. 
Within the manuscript we denote the mean rupture time 
by $\tau(F)$ with 
\be 
\tau(F)=T_1(b_r,b=b_1), 
\ee to highlight its
dependency on $F$, because $b_r$ is a bond type-specific constant, 
because we are not considering higher moments than the first moment, 
and because we choose the bond to reside at $t=0$ in its energetic minimum,
located at $b=b_1(F)$. 
Recall that our potential has the features $U(b_1)=U'(b_1)=0$
and $U''(b_1)>0$.
Moreover, we use $b_0=0$ as the reflecting boundary, noting that the precise
choice does not matter as $U$ tends to diverge at $b\rightarrow 0$ due to excluded
volume interactions.
We checked that the $\tau(F)$ calculated semi-analytically via Eq.\ \eqref{finalT1} 
with Eq.\ \eqref{finalPsi} 
(numerical integration of the double-integral) is exactly identical with
the mean rupture time obtained via Brownian dynamics of the Langevin equation \eqref{Langevinbond} with reflecting boundary at $b_0=0$, adsorbing boundary at $b_r$,
and initial condition $b(0)=b_1(F)$, 
if results are extrapolated to infinitely small time step. 


\subsection{Influence of boundary conditions}

Explanation of \textbf{fix-move}, \textbf{fix-force}, \textbf{fix-smd} and \textbf{fix-spring}.
fix-smd is the primary method for chain stretching used within the manuscript. 
Only Figure~1(c) is obtained using fix-move and fix-force, while all other figures are obtained with fix-smd.\\

\textbf{fix-move}: move the positions of 2 terminal Si atoms of a PDMS chain with constant velocity $V$ (i.e. stretch a chain with constant velocity $V$). When $V$ is set to 0, we can fix the extension $R$ of a PDMS chain and measure the force $F$ to obtain force-extension relation.

\textbf{fix-force}: apply a constant force $F$ on a terminal Si atom while fixing the position of the other terminal Si atom (i.e. stretch a polymer chain with constant force $F$). After equilibrium is reached, we register $R$ to measure force extension.

\textbf{fix-smd}: apply a constant repulsive force $F$ between 2 terminal Si atoms. After equilibrium is reached, we register $R$ to measure force extension. 

\textbf{fix-spring}: use 2 springs to stretch the 2 terminal Si atoms of a PDMS chain. We can measure the $R$ and $F$ (forces in 2 springs) to obtain force-extension. However due to the oscillation of springs, the measurement error is extremely large so no results presented in this manuscript are measured with this method.

\begin{figure*}
    \centering
    \includegraphics[width=0.8\textwidth]{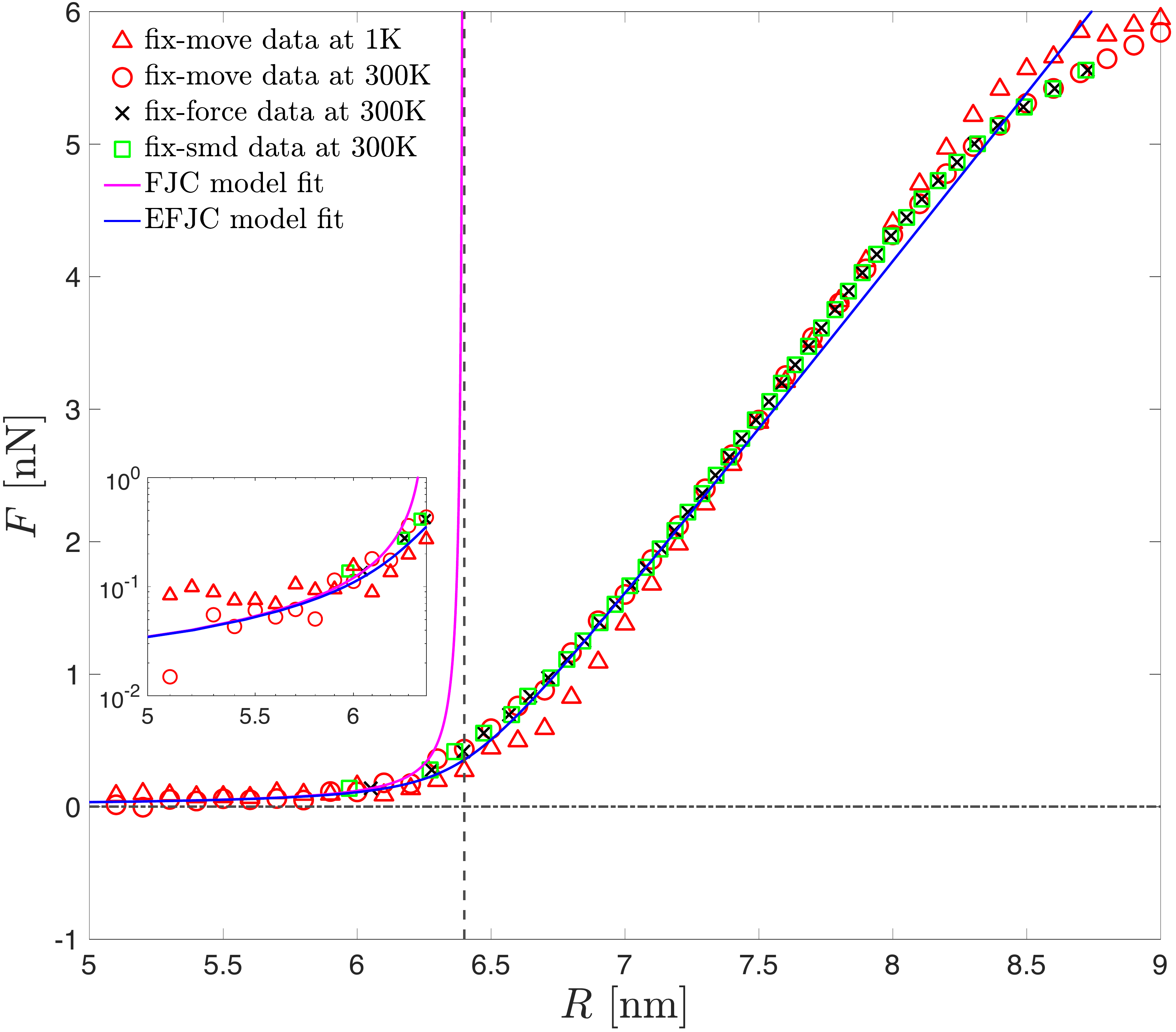}
    \caption{Investigation of the influence of boundary conditions, temperature, and the presence/absence of solvent molecules on the force-extension relation of PDMS-27.
    We find that the choice of boundary condition does not influence the obtained force-extension relation (fix-spring is omitted, as it is not suitable in force-extension measurements due to large measurement errors linked to spring oscillations). fix-move measurements at $T=$ 1K (red triangles) show a slightly smoothed-out transition between entropic- and enthalpic regimes, which stems from reduced entropic contributions.}
    \label{fig-s2}
\end{figure*}

\begin{figure*}
    \centering
    \includegraphics[width=0.7\textwidth]{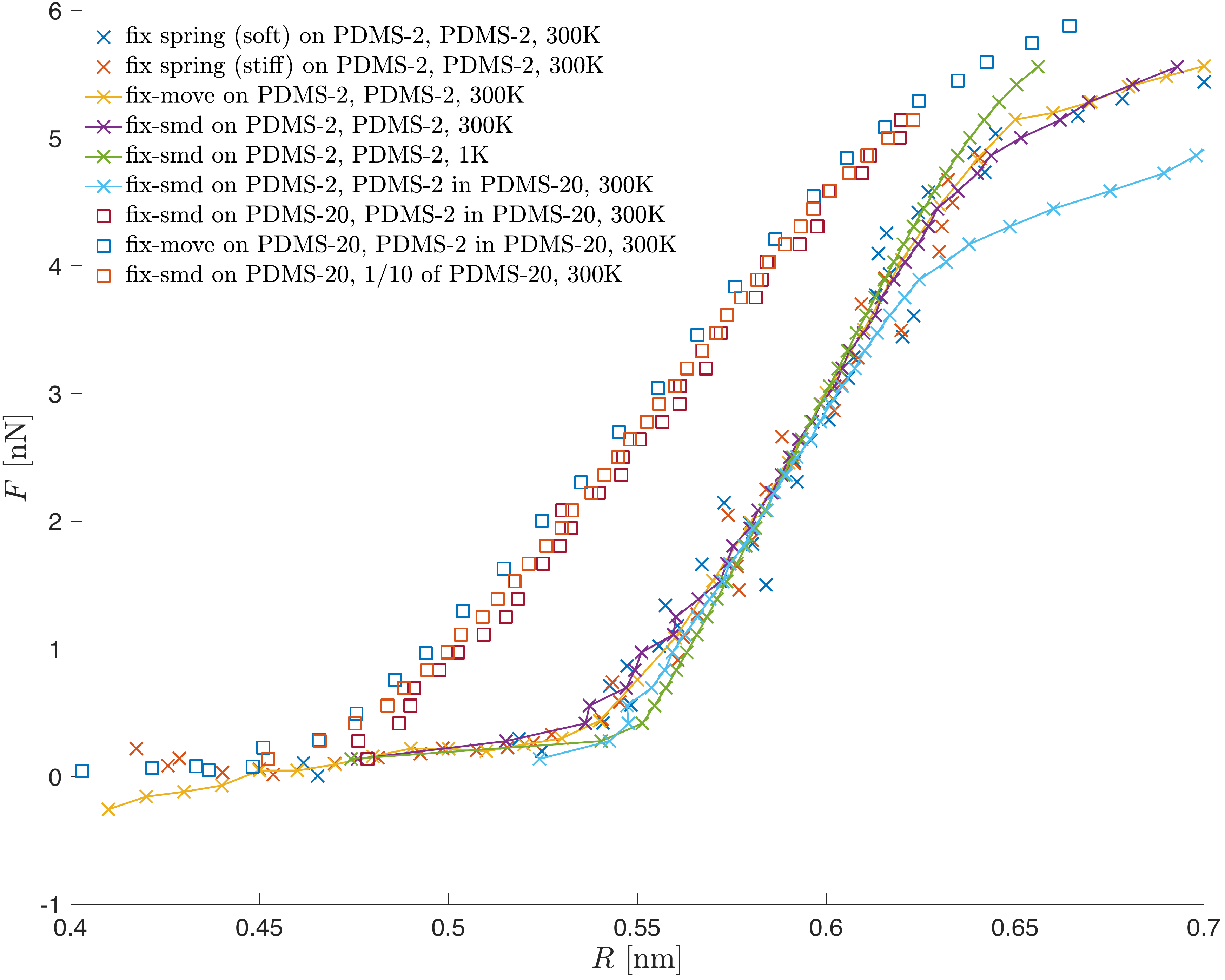}
    \caption{Investigation of the influence of boundary conditions on the observed 'end effect'.
    We focus on three different molecules: PDMS-2, PDMS-20, as well as a PDMS-2 segment (Si-Si-Si) located at the center of PDMS-20. 
    Different boundary conditions do not influence the force-extension relation within the enthalpic part.
    Furthermore, stretching an isolated PDMS-2 molecule is equivalent of stretching PDMS-2 at the center of PDMS-20.
    The difference between data plottes as $\times$ and $\square$ is the chain length between the 2 stretching points.}
    \label{fig-s4}
\end{figure*}

\newpage

\subsection{Bond length distributions and potentials}

\begin{figure*}
    \centering
    (\textbf{a})\includegraphics[width=0.3\textwidth]{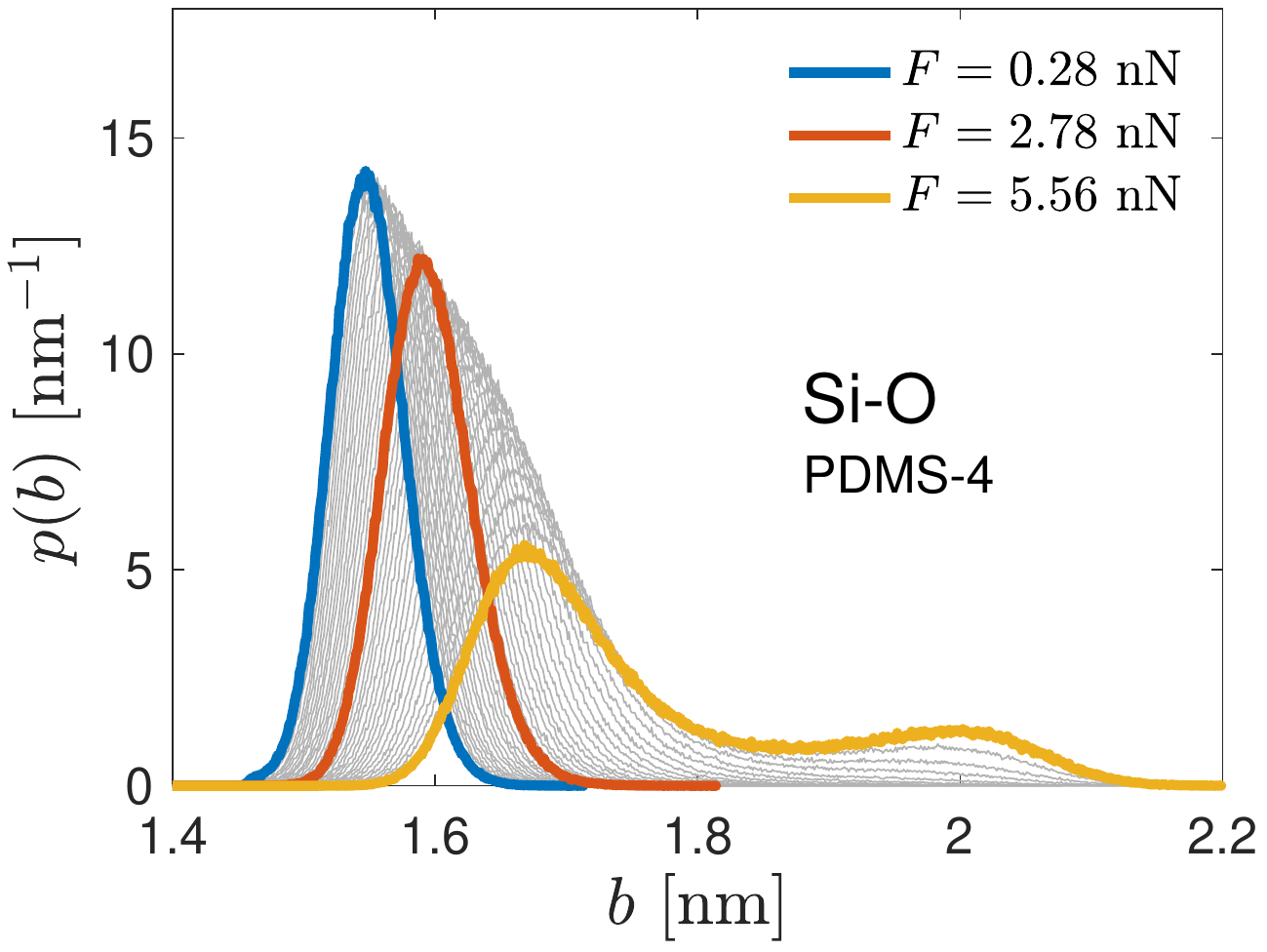}
    (\textbf{b})\includegraphics[width=0.3\textwidth]{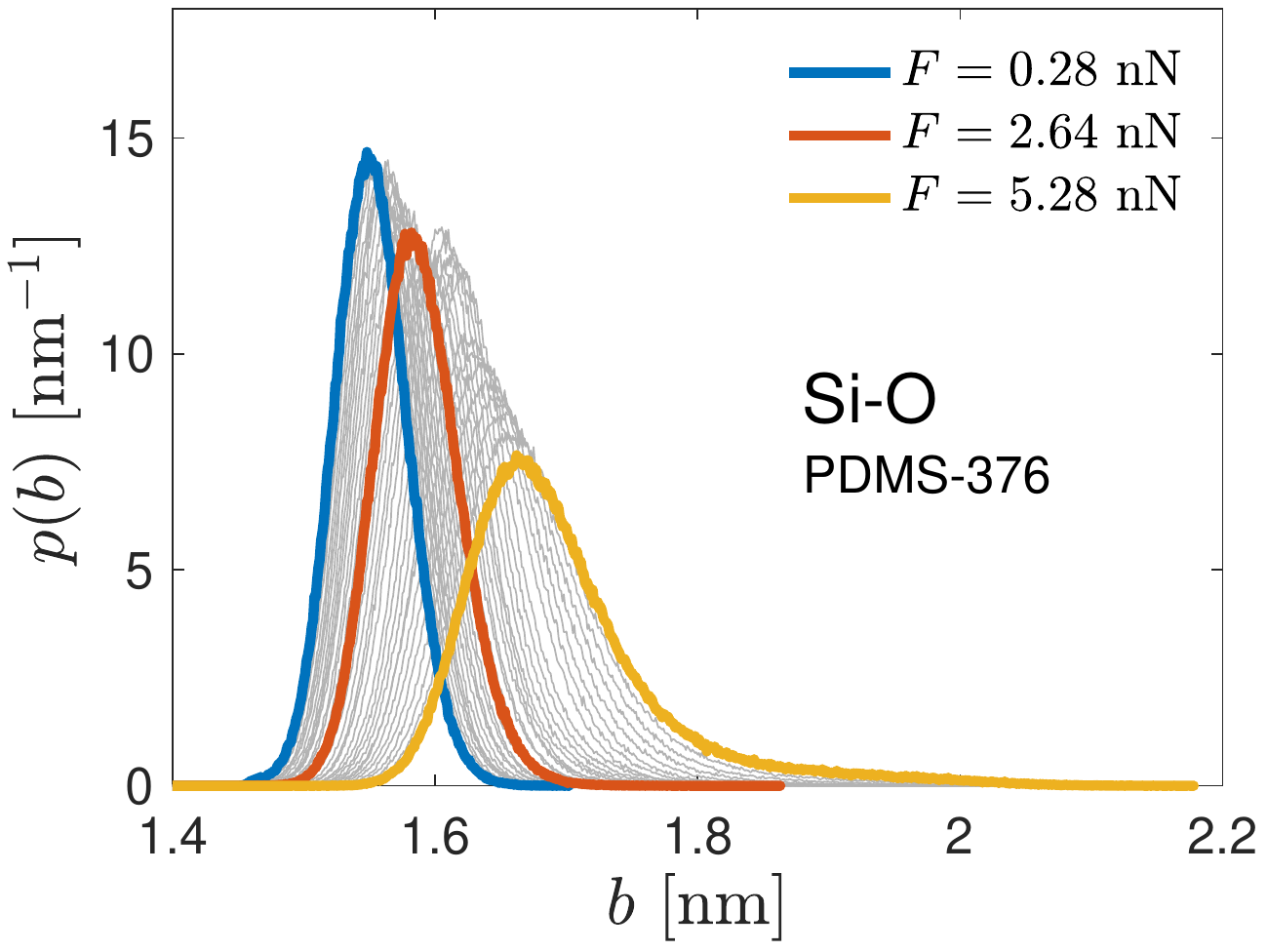}\\
    (\textbf{c})\includegraphics[width=0.3\textwidth]{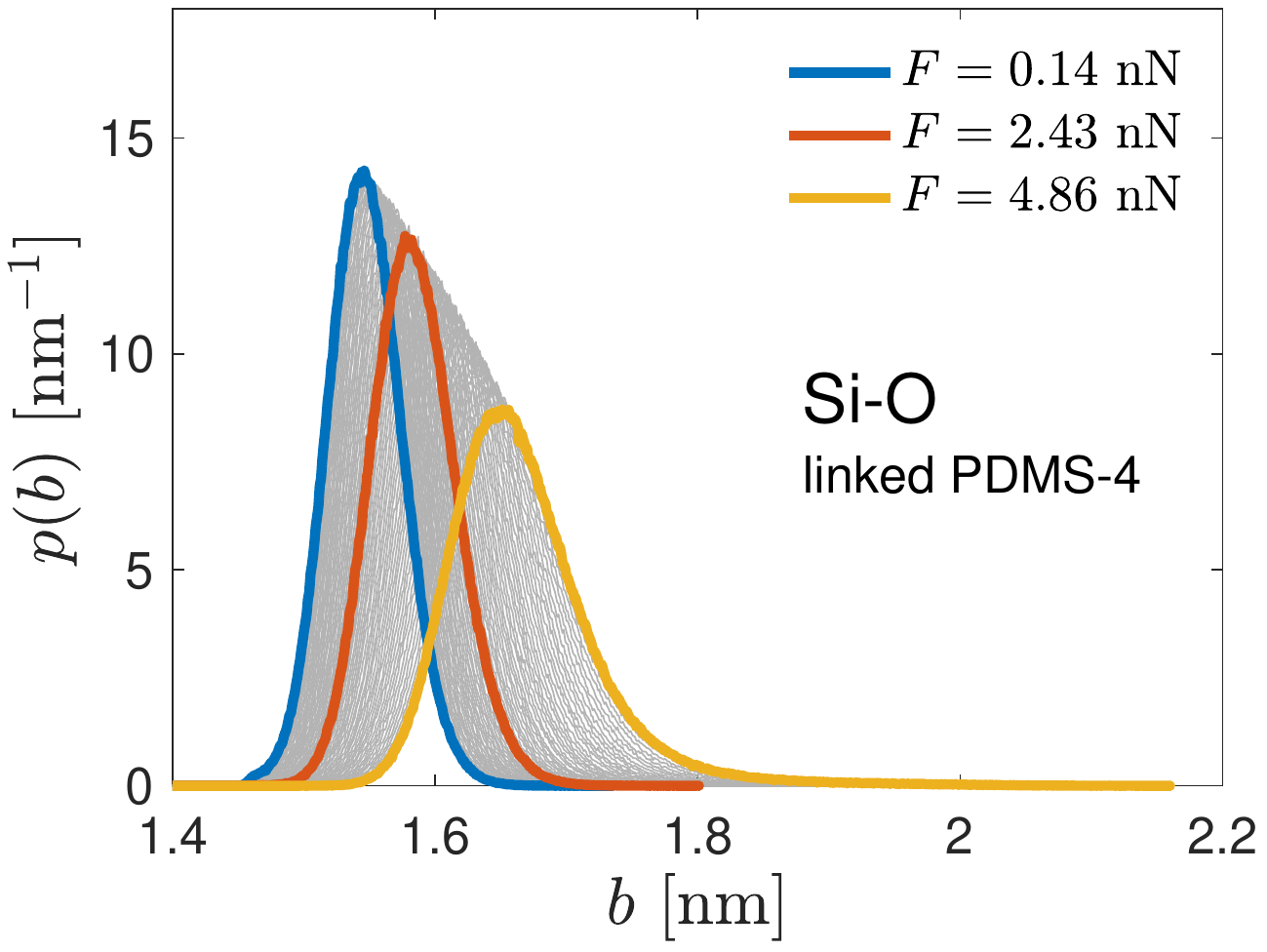}
    (\textbf{d})\includegraphics[width=0.3\textwidth]{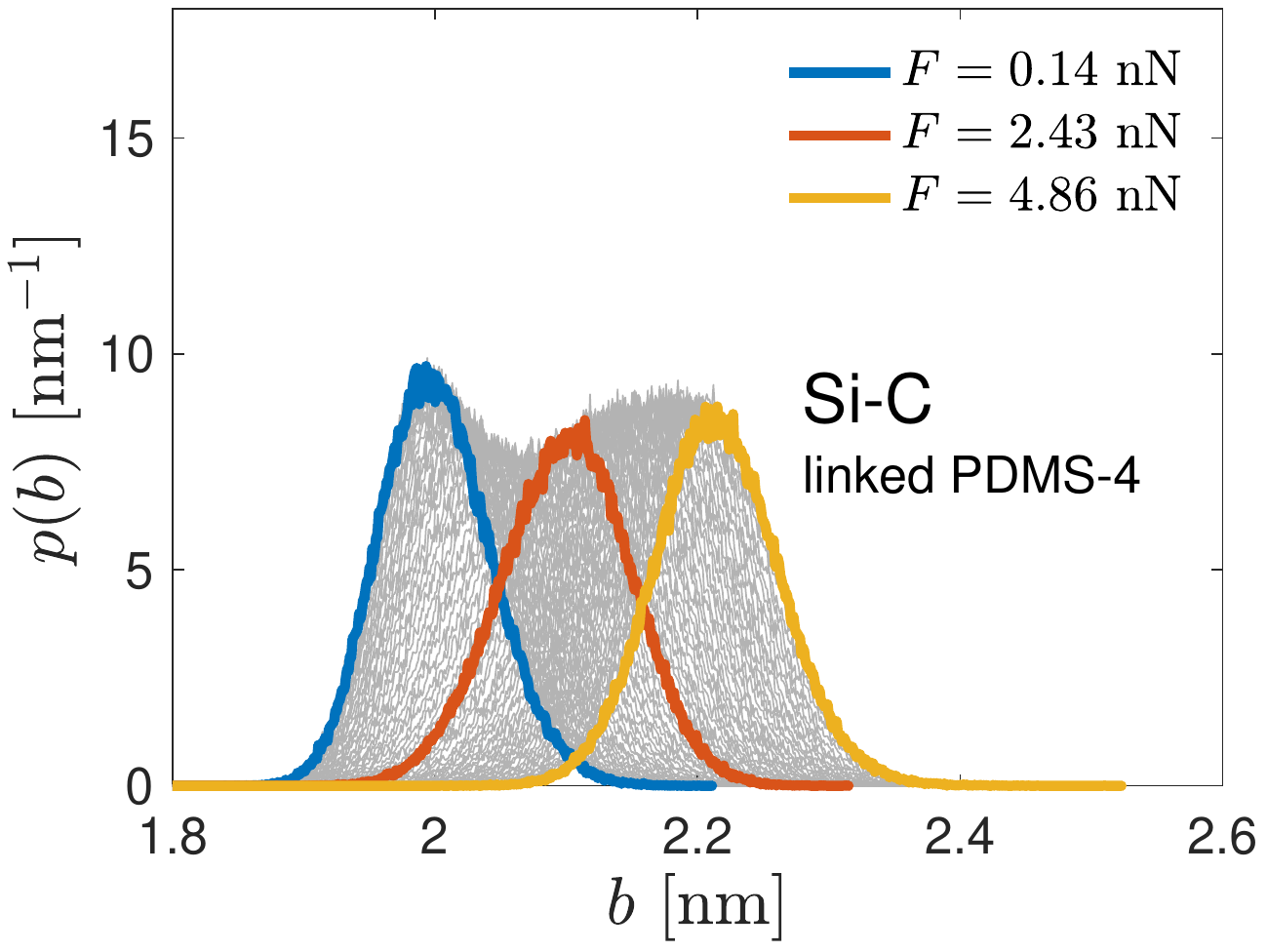}
    (\textbf{e})\includegraphics[width=0.3\textwidth]{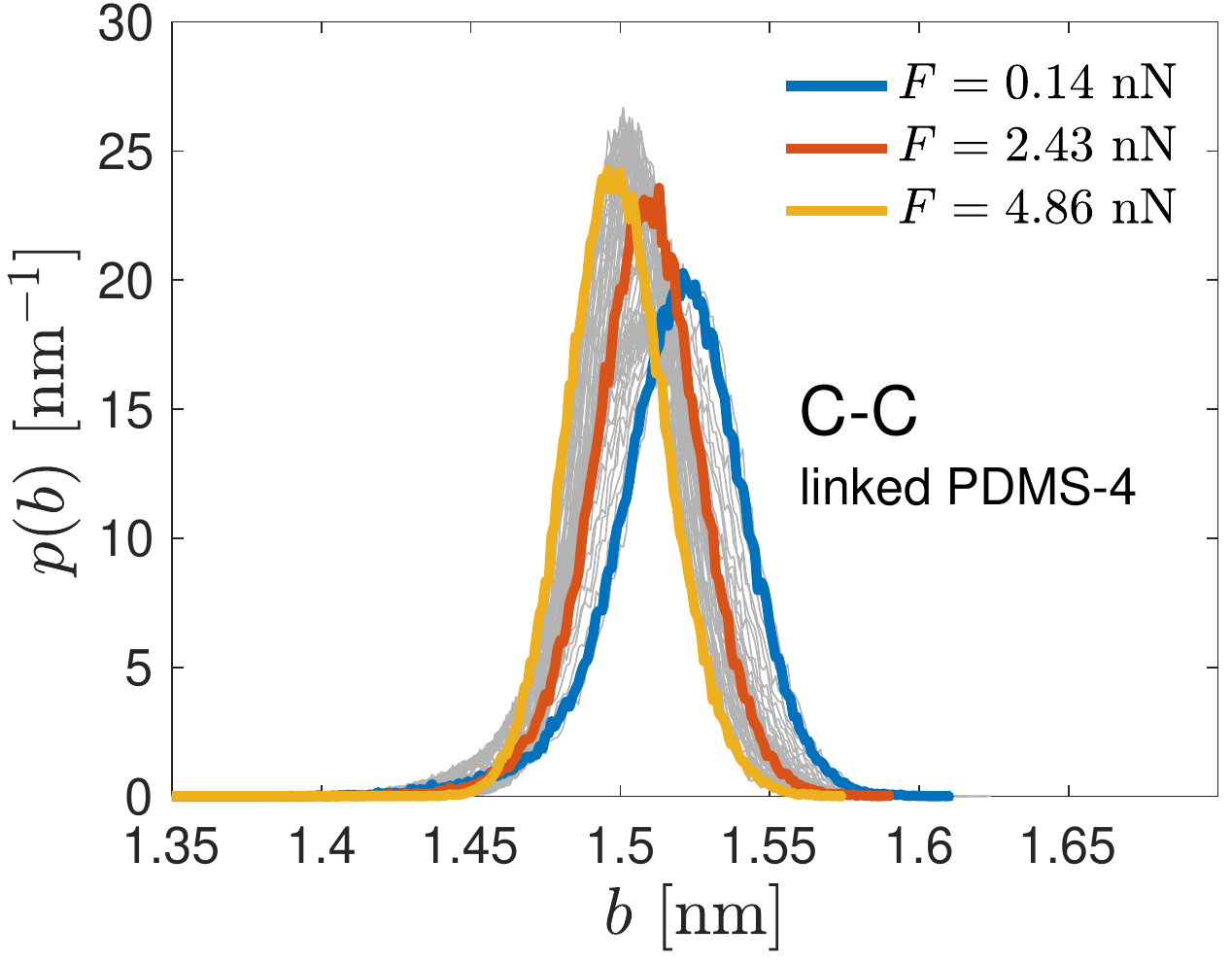}
    \caption{Bond length distribution on (\textbf{a}) PDMS-4, (\textbf{b}) PDMS-376 and (\textbf{c-e}) linked PDMS-4
    obtained from atomistic MD. Simulations are performed at $T=300$ K.}
    \label{fig-s5}
\end{figure*}

\begin{figure*}
    \centering
    (\textbf{a})\includegraphics[width=0.3\textwidth]{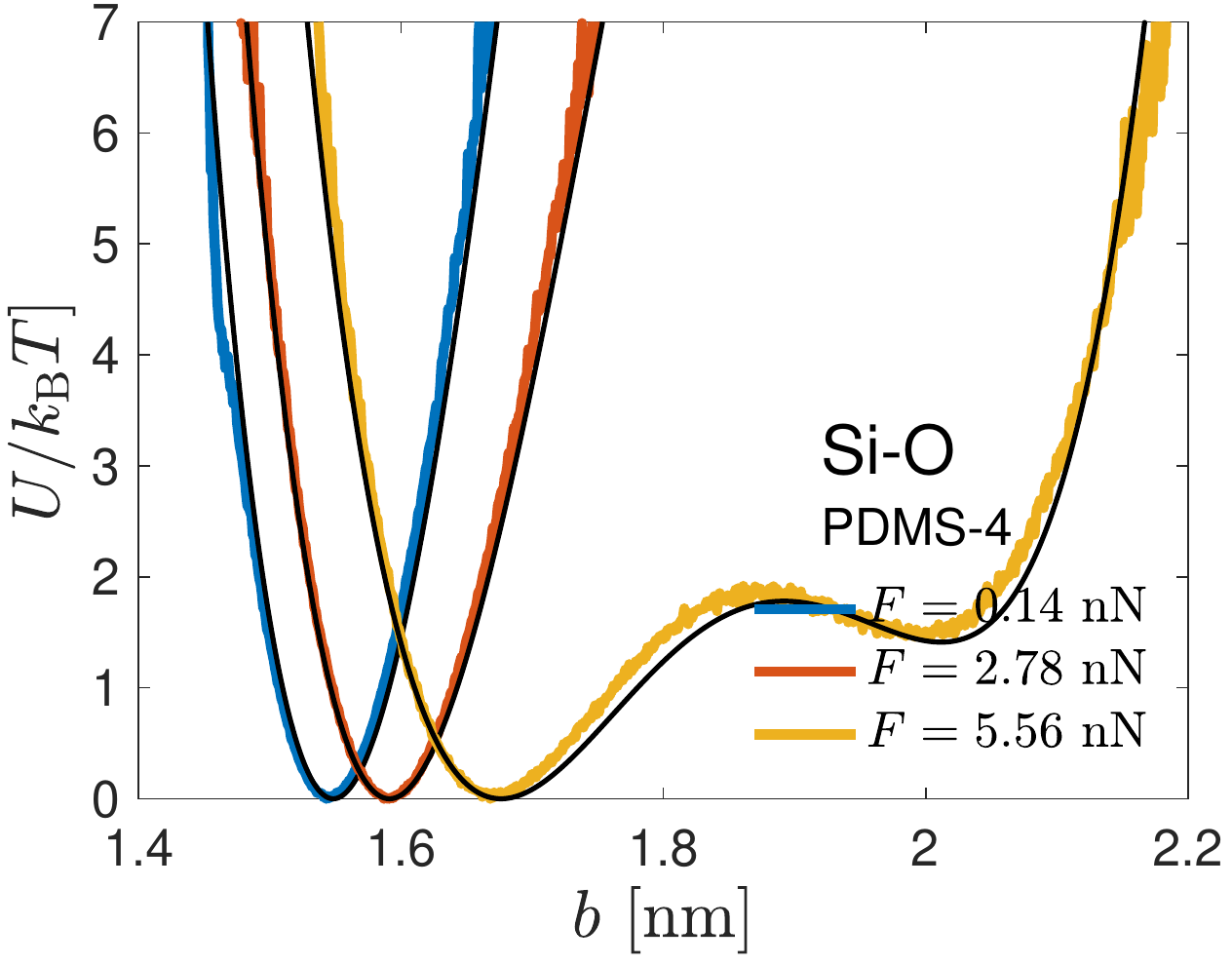}
    (\textbf{b})\includegraphics[width=0.3\textwidth]{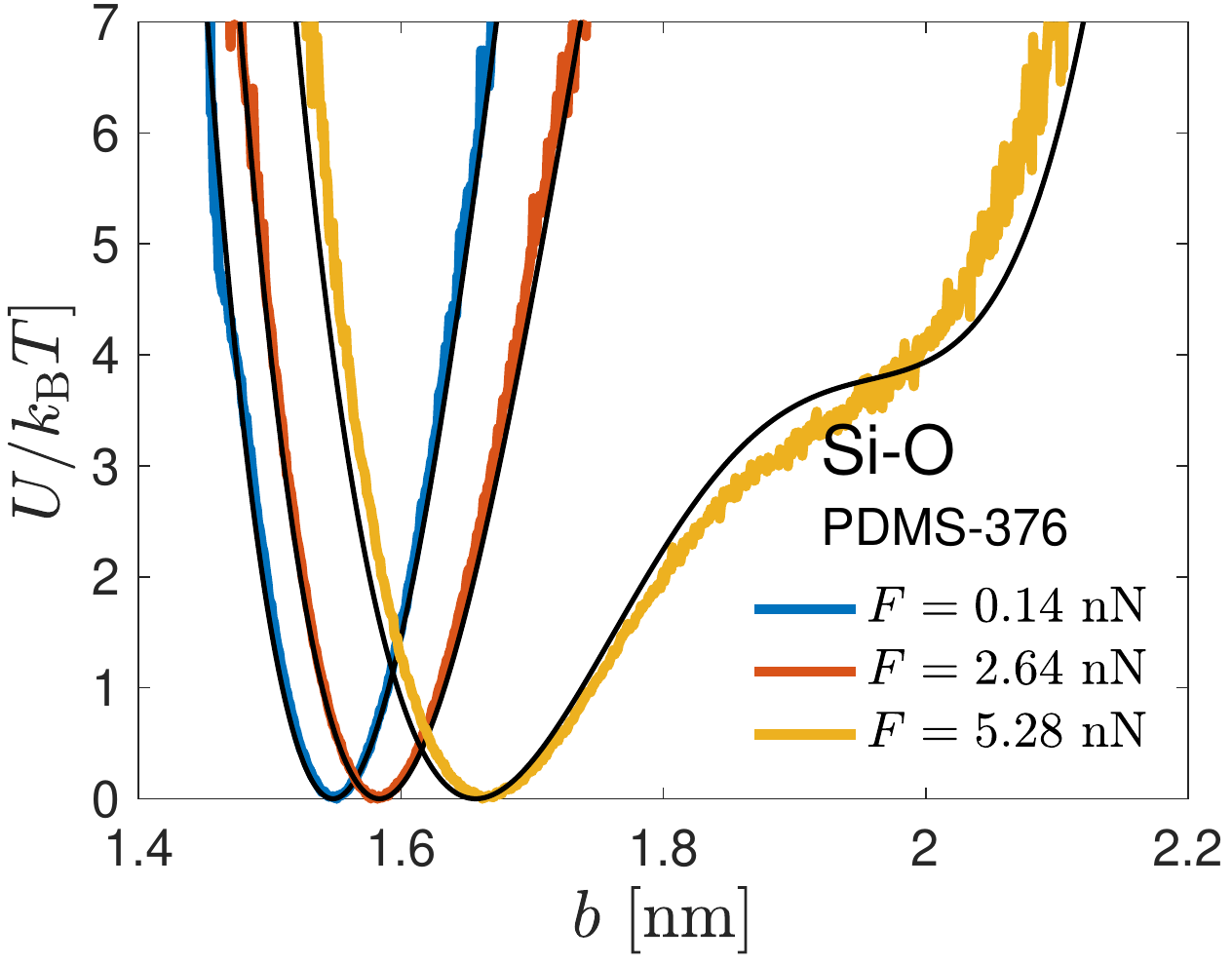}\\
    (\textbf{c})\includegraphics[width=0.3\textwidth]{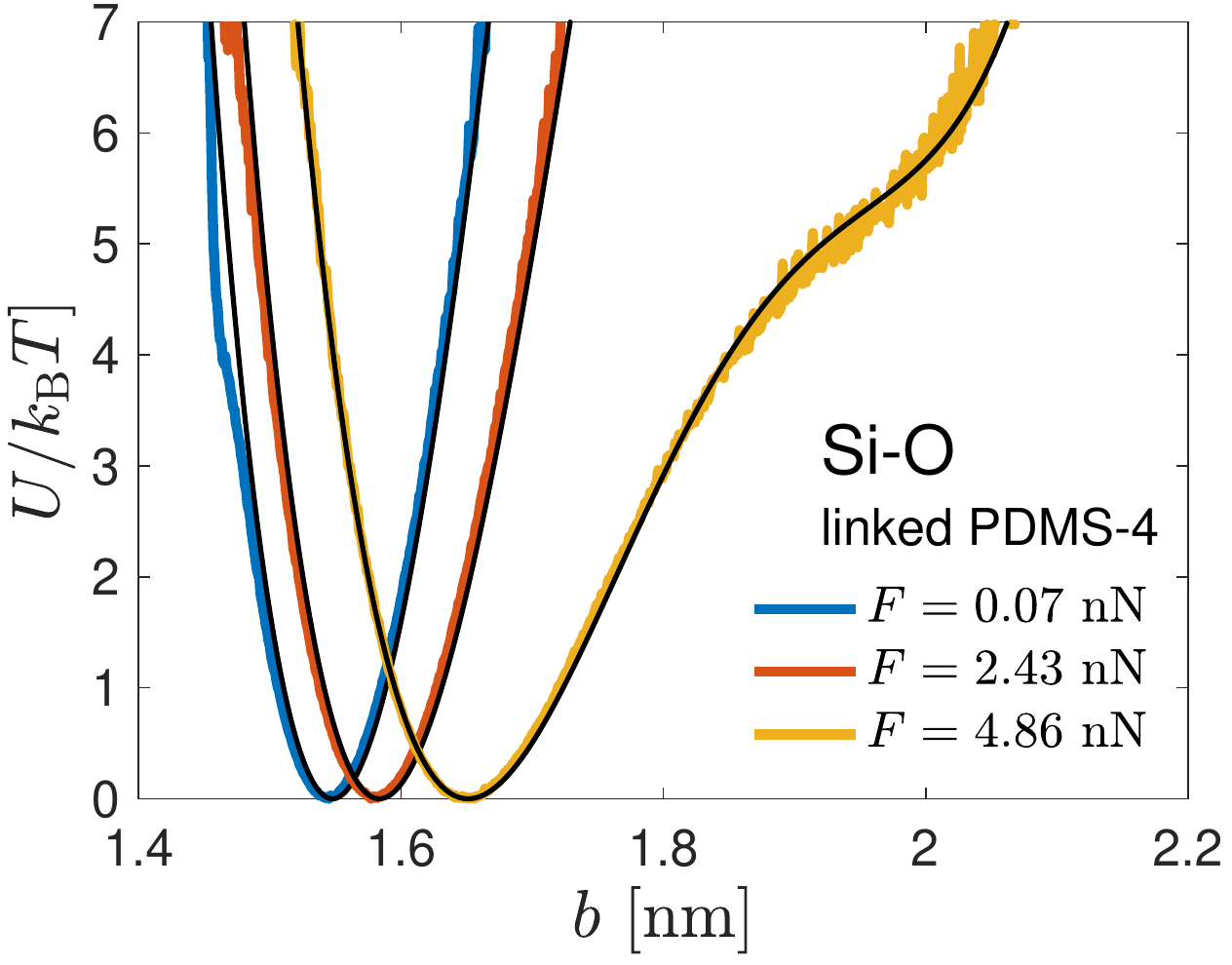}
    (\textbf{d})\includegraphics[width=0.3\textwidth]{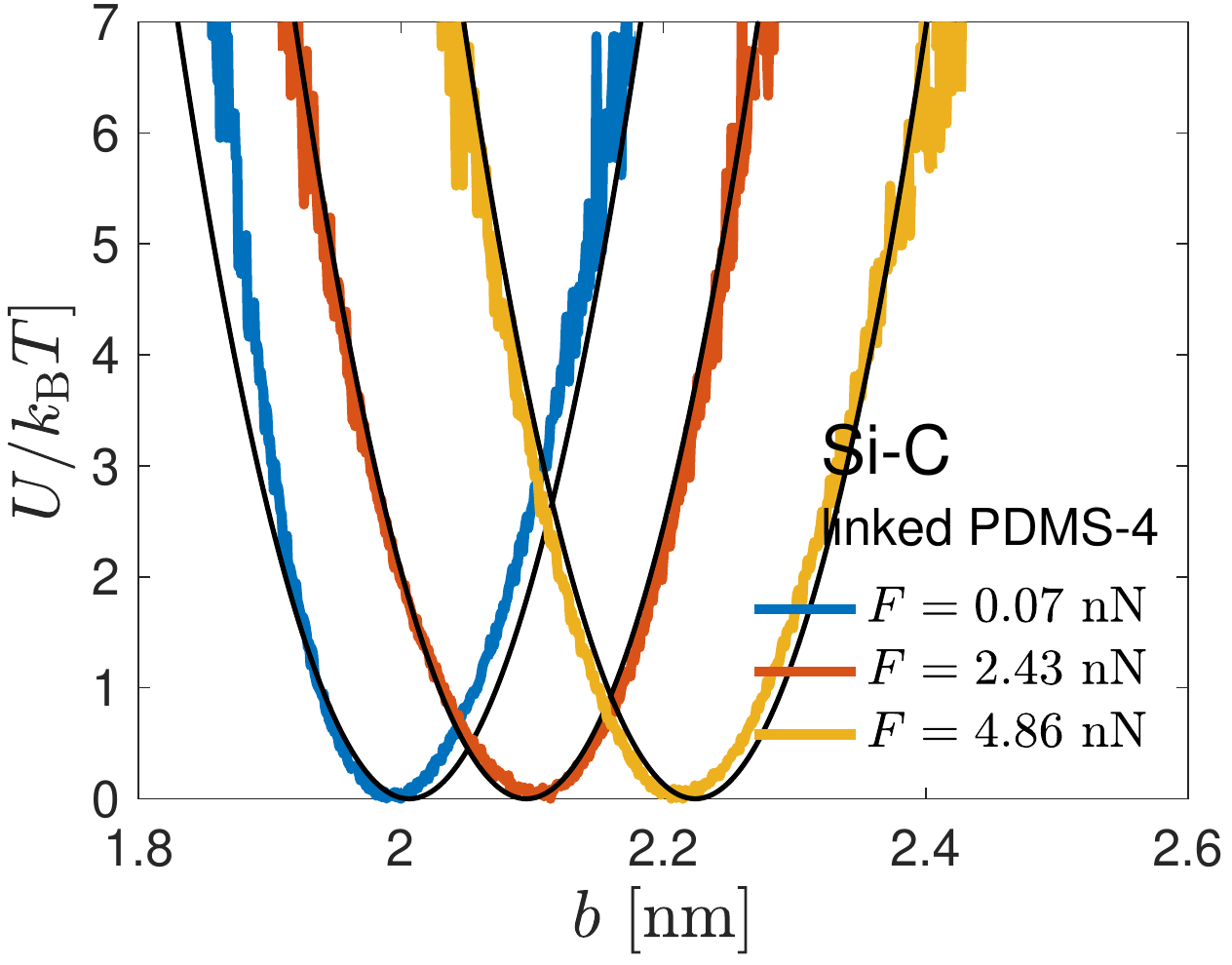}
    (\textbf{e})\includegraphics[width=0.3\textwidth]{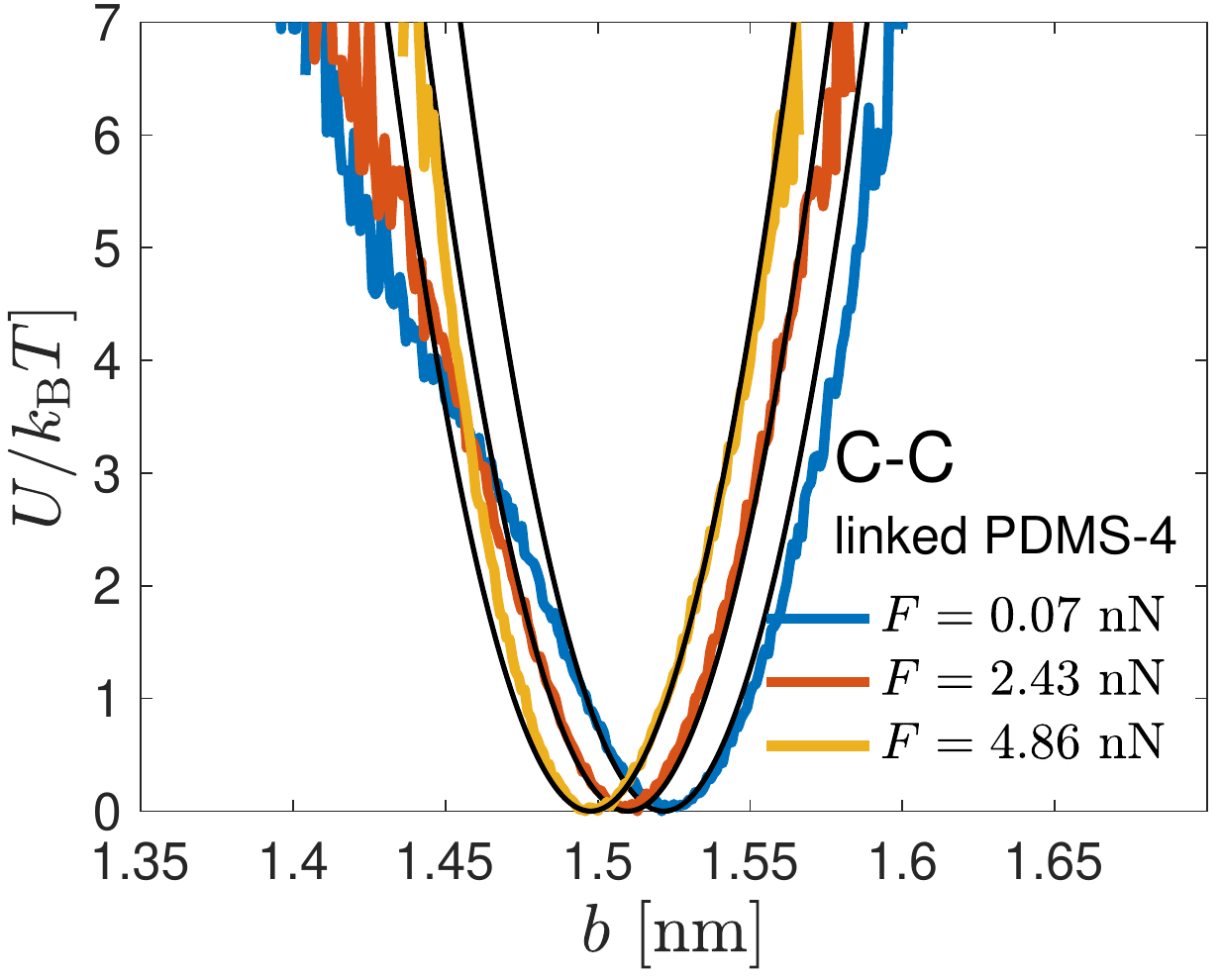}
    \caption{Effective bond potentials calculated from the bond length distributions on (\textbf{a}) PDMS-4, (\textbf{b}) PDMS-376 and (\textbf{c-e}) linked PDMS-4, shown in Fig.\ \figref{fig-s1}, along with the Si-O and Si-C fit functions (solid black lines) stated in the manuscript.}
    \label{fig-s6}
\end{figure*}

\begin{figure*}
    \centering
    \includegraphics[width=\textwidth]{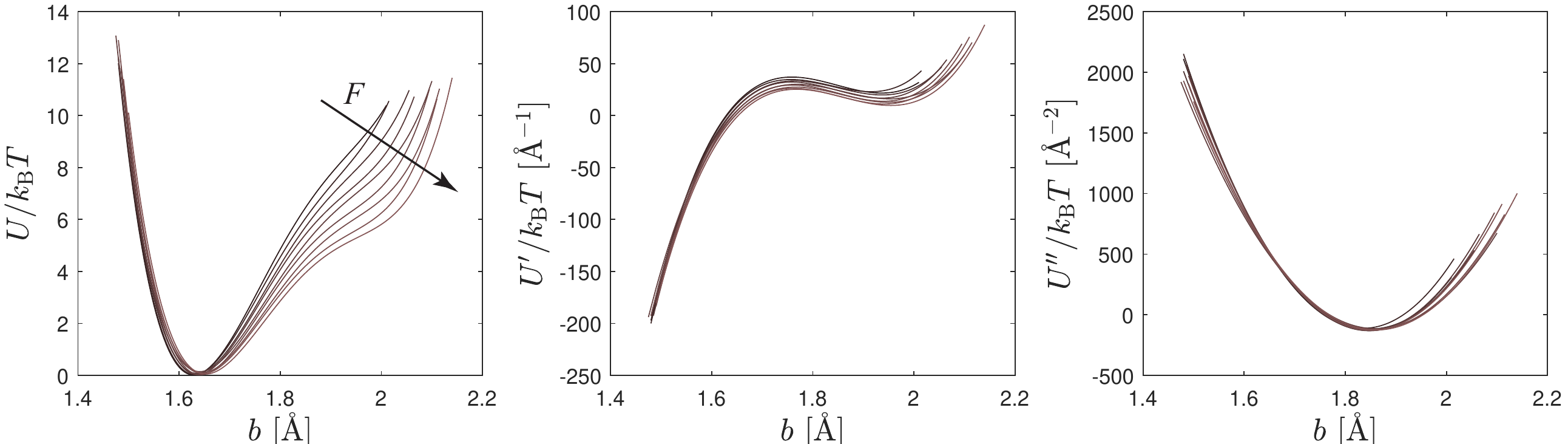}
    \caption{Potential of mean force and its derivatives, $U(b)$, $U'(b)$ and $U''(b)$ for Si-O at different levels of relatively strong applied force $F>3$ nN (MD simulations performed in the presence of solvent molecules). Since $U''$ exhibits parabolic
    and $F$-independent shape and location, the corresponding parameters $b_2$, $c_2$, $k_2$ in the 4th order polynomial are treated as $F$-independent constants.}
    \label{fig:si_potential_derivs}
\end{figure*}

\clearpage
\subsection{Rupture times}

\begin{figure*}
    \centering
    \includegraphics[width=0.7\textwidth]{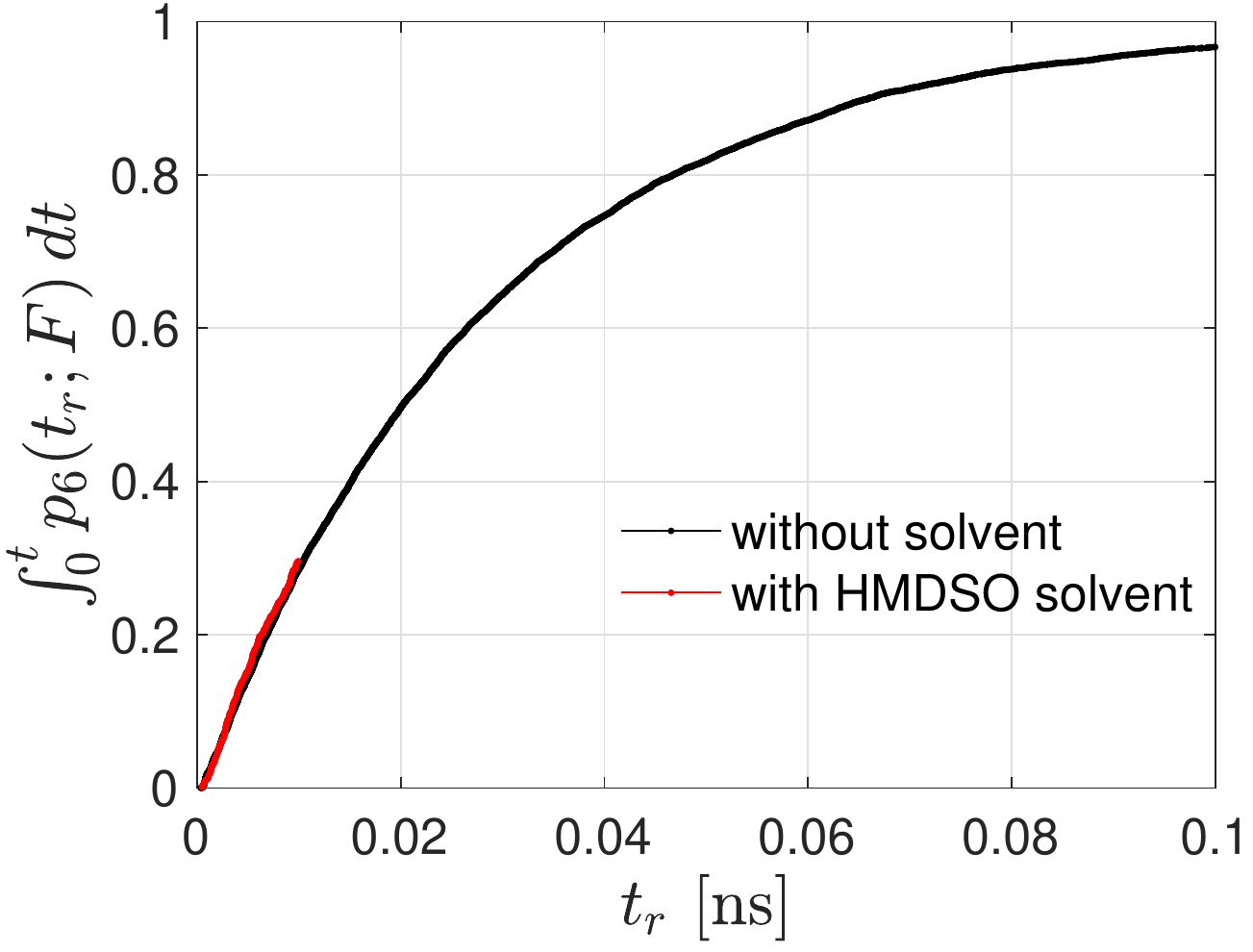}
    \caption{Cumulative rupture time distribution for PDMS-6 at $F=6.39$ nN. Red: Simulation in the presence of HMDSO solvent molecules. Black: Simulations without solvent.
    Results obtained by averaging over 100 (with) and 10000 (without solvent) independent start configurations. Simulations are performed at $T=300$ K.}
    \label{fig:rupture}
\end{figure*}

\begin{figure*}
    \centering
    \includegraphics[width=0.85\textwidth]{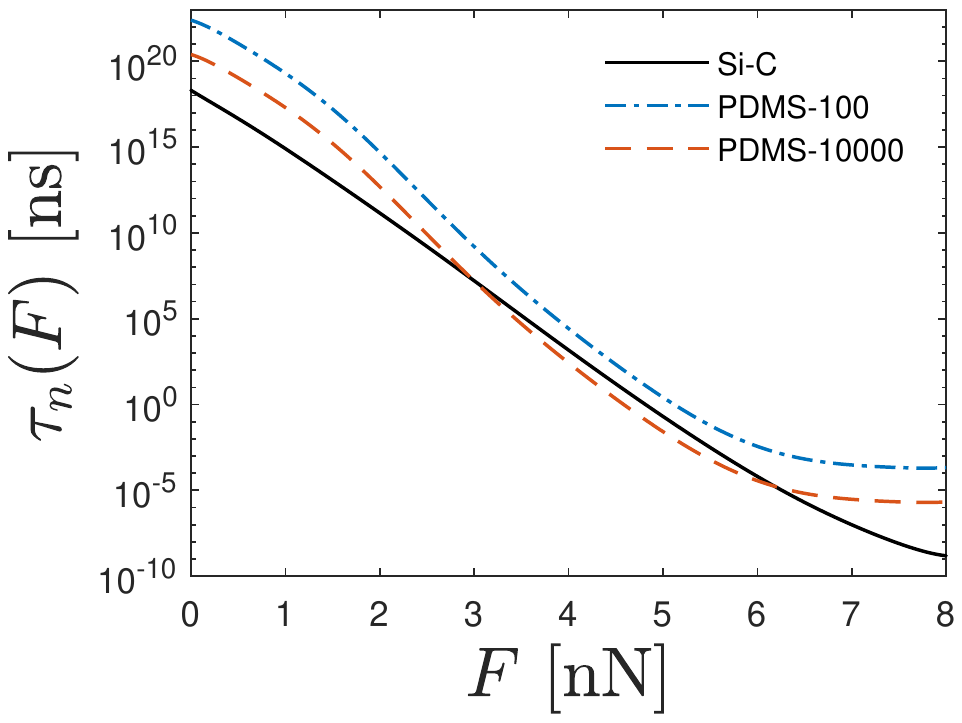}
    \caption{Equivalent of Figure~\figref{fig3}c for different polymerization degrees $n$. At low $n$ (blue dashed line), crosslinking junctions are weaker than Si-O bonds along the backbone (corresponding to the dashed line at position (A) in Figure~\figref{fig4}c. A double crossover between Si-C and Si-O bonds is observed at high $n$ (dashed red line). This corresponds the re-entrant effect observed in Figure~\figref{fig4}c.}
    \label{fig-4c}
\end{figure*}

Here, we provide evidence that the measured rupture times are basically unaffected by the presence of HMDSO solvent molecules. 
In the absence of solvent, the friction coefficient $\zeta$ is implicitly captured by the employed thermostat. 
This finding allows us to simulate the exponential tail of the rupture time distribution in the absence of solvent (Fig.\ \figref{fig3}a).
This renders computatiosn feasible (as it is two orders of magnitude cheaper than the full atomistic simulation of the solvated PDMS chain). 
Shown in Fig.~\figref{fig:rupture} is the cumulative fraction of ruptured PDMS-6 chains versus time both in the presence and absence of solvent molecules.

\newpage

\subsection{Mean chain rupture times for different polymerization degrees}

\end{document}